\documentclass[11pt]{article} 

\usepackage[utf8]{inputenc}
\usepackage{enumitem} 
\usepackage[margin=2.5cm]{geometry} 
\usepackage[normalem]{ulem} 
\usepackage{fancyhdr} 
\usepackage[english]{babel}

\usepackage{amsmath, amssymb} 
\usepackage{esvect} 
\usepackage{physics} 
\usepackage{mathtools} 
\usepackage{textcomp} 
\usepackage{gensymb} 
\usepackage[T1]{fontenc} 
\usepackage{bm} 
\usepackage{dsfont} 

\usepackage{setspace} 

\usepackage{graphicx} 
\usepackage{caption} 
\usepackage{subcaption}
\usepackage[labelsep=period]{caption} 
\usepackage{tikz,lipsum,lmodern} 
\usepackage{tcolorbox} 
\usepackage{float} 
\usepackage{sidecap} 

\usepackage[backend=biber,style=nature]{biblatex}
\title{One-dimensional Lieb superlattices: from the discrete to the continuum limit} 
\author{Dylan Jones$^1$, Marcin Mucha-Kruczynski$^1$, Adelina Ilie$^1$\footnote{Corresponding Author: A.Ilie@bath.ac.uk}, Lucian Covaci$^{2,3}$\footnote{Corresponding Author: lucian.covaci@uantwerpen.be}\\
$^1$ Department of Physics, University of Bath, Bath, UK\\
$^2$ Department of Physics, University of Antwerp, Antwerp, Belgium\\
$^3$ NANOlab Center of Excellence, University of Antwerp, Antwerp, Belgium} 
\date{\today} 
\begin{document} 
\maketitle 
\begin{abstract} 
The Lieb lattice is one of the simplest lattices that exhibits both linear Dirac-like and flat topological electronic bands. We propose to further tailor its electronic properties through periodic 1D electrostatic superlattices (SLs), which, in the long wavelength limit, were predicted to give rise to novel transport signatures, such as the omnidirectional super-Klein tunnelling (SKT). By numerically modelling the electronic structure at tight-binding level, we uncover the evolution of the Lieb SL band structure from the discrete all the way to the continuum regime and build a comprehensive picture of the Lieb lattice under 1D potentials. This approach allows us to also take into consideration the discrete lattice symmetry-breaking that occurs at the well/barrier interfaces created by the 1D SL, whose consequences cannot be explored using the previous low energy and long wavelength approaches. We find novel features in the band structure, among which are intersections of quadratic and flat bands, tilted Dirac cones, or series of additional anisotropic Dirac cones at energies where the SKT is predicted. Such features are relevant to experimental realizations of electronic transport in Lieb 1D SL realized in artificial lattices or in real material systems like 2D covalent organic/metal-organic frameworks and inorganic 2D solids. 
\end{abstract} 
%
%
\section*{Introduction}
Systems that exhibit non-trivial band topology and flat bands have become the focus of intense recent research. 
One such system is the Lieb lattice, a two-dimensional (2D) depleted-square lattice that supports a graphene-like conical dispersion intersected by a flat band at zero energy. 
Although models exhibiting flat bands were originally seen as theoretical toy models, recent experimental developments in ultra cold gases \cite{taie2015coherent}, photonics \cite{vicencio2015observation, huang2011dirac, guzman2014experimental} or artificial lattices \cite{moitra2013realization} make them a reality. 
Associated with a large enhancement of the density of states due to quenching of the kinetic energy, flat bands in the Lieb lattice are expected to host many-body phenomena such as superconductivity \cite{huhtinen2022revisiting}, ferromagnetism \cite{jiang_lieb-like_2019} and topologically non-trivial states \cite{jiang2020topological}. 
A well known example is the ``magic'' angle twisted bilayer graphene, which showcases the importance of band flatness and its topology in the appearance of correlated states \cite{cao2018unconventional}. 
Similarly, it was shown that the flat band realised in the Lieb lattice can sustain robust superconductivity due to its non-trivial quantum metric \cite{julku2016geometric, huhtinen2022revisiting}.
%

Experimental and theoretical works have shown that the electronic properties of electronic systems can be tailored by the application of periodic potentials.  
For example, when a one-dimensional periodic (1D) potential is applied to graphene, additional Dirac cones are created, flattened in the direction perpendicular to the direction of the applied potential, enabling tunable anisotropy of the group velocity \cite{park2008anisotropic, barbier2010extra, li2021anisotropic}. 
Such periodic potentials can be applied through top-down patterned gating, with wavelength as low as $50$ nm \cite{li2021anisotropic} or through bottom-up self-assembly of macromolecules on top of van der Waals materials, with wavelengths as low as 3-5 nm  \cite{gobbi2017periodic, tsai2020molecular, gobbi2018collective}. 

Analytical models predict the existence of super-Klein tunnelling (SKT) phenomena in the Lieb lattice (and other pseudospin-1 systems) under a 1D periodic potential \cite{xu2014omnidirectional, cunha2021band, betancur-ocampo_super-klein_2017}. 
Different to the Klein tunneling in the graphene lattice (pseudospin-1/2) \cite{katsnelson2006chiral}, the SKT state manifests as unity transmission probability irrespective of the angle of incidence \cite{urban2011barrier, illes2017klein, fang_klein_2016, shen_single_2010}. 
This was recently claimed to be demonstrated in a triangular phononic crystal \cite{zhu2023experimental} by investigating transport across a singular potential barrier. 
However, these predictions came with two important caveats: (i) the potential's lengthscale is very large compared to the lattice constant (i.e. the long wavelength or the continuum limit), and (ii) these systems do not require beyond nearest-neighbour interactions to be described accurately.  This limits the applicability of the analytical predictions to a wider range of experimentally realisable systems. 

The simplest approach towards describing the peculiar electronic properties of the Lieb lattice is based on a tight-binding model with only nearest-neighbour site interactions. However, both in artificial lattices and in a newer class of solid-state materials that can possess a Lieb lattice structure, such as recently proposed covalent-organic and metal-organic frameworks (COFs and MOFs), the presence of additional next-nearest neighbour (NNN) hoppings disperses the flat band and lattice distortions can easily give rise to band gaps \cite{cui2020realization, jiang2020topological, jin2017two}. 
Additionally, a finite spin-orbit interaction, introduced by the presence of metallic atoms will give rise to topological gaps in the band structure, resulting in topological spin-Hall edge states that are protected from back-scattering \cite{goldman_topological_2011}.  
\textit{It is expected that these additional effects will strongly influence  many-body correlations predicted for the Lieb lattice flat bands}.

In light of growing interest in experimental realizations of the Lieb lattice, further tunability through (1D) periodic potentials will become possible in the near future. 
Nevertheless, a general description of the electronic properties of the Lieb lattice under a periodic potential landscape is lacking. 
Important aspects related to the lattice discreetness, band dispersion and distortion-induced gaps remain largely unexplored. 
Questions whether the universal nature of the SKT survives arise when the discreteness of the lattice has to be taken into account.

Here, we numerically model the electronic structure of a Lieb lattice under a 1D periodic potential (1D Lieb superlattice). This approach is general, applicable to the description of a range of systems from trapped cold atoms and artificial Lieb lattices to solid-state realisations like recently synthesised (M-)COFs, from the discrete (where the potential periodicity is on the order of the lattice constant) to the continuum limit. Using a tight-binding (TB) model permits the inclusion of next-nearest neighbour interactions, an effective mass term, and spin-orbit coupling, required to describe experimentally relevant realisations of the Lieb lattice. Additionally, a tight-binding description provides insights into discrete lattice symmetry-breaking effects at the well/barrier interfaces imposed by the periodic potential and allows the identification of non-trivial topological properties of the system. 

We find that in both limits, i.e. the discrete ($L\sim a$) and the continuum ($L\gg a$) limits, where $L$ is the superlattice wavelength and $a$ is the Lieb lattice constant, the discreteness of the lattice plays an important role in the resulting energy dispersion of the Lieb 1D superlattice. For example, tilted Dirac cones, intersections of quadratic and flat bands and new Dirac cones are found in the discrete limit. In the continuum limit, when comparing to low-energy and long-wavelength approximations near the Dirac cone, although propagating states are recovered, the SKT state is absent in the tight-binding simulations. Instead, we uncover a sequence of anisotropic Dirac cones, reminiscent of the behaviour of graphene under a periodic 1D potential landscape showing a sequence of Dirac cones with frequency dependent on the ratio $L/a$. Furthermore, we find that a smooth interface gives rise to further localized states that interfere with the propagating states, while the SKT states are still absent, replaced by dispersive anisotropic Dirac cones. We give a special attention to more complex Lieb lattice models that take into account next-nearest neighbor (leading to dispersive bands), imbalance in the on-site energies in the unit cell (leading to band gap opening at the Dirac point) and spin-orbit coupling terms (leading to topological band gaps). With the addition of these terms, corresponding to realistic and experimentally relevant realizations of the Lieb lattice, we provide a comprehensive picture on their effect on the electronic properties of realistic 1D Lieb superlattices.
%
%
%
%
\section*{The discrete limit}
\begin{figure}[H]
    \centering
    \includegraphics[width=\textwidth]{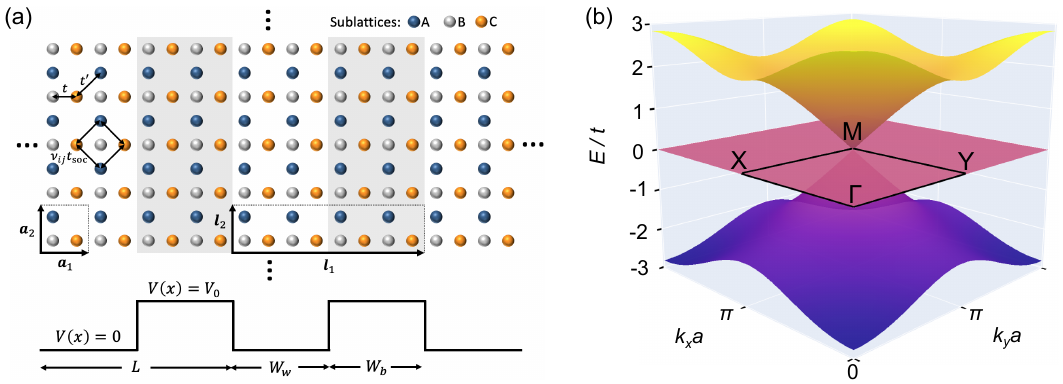}
    \caption{\textbf{Lieb superlattice (SL) model diagram and pristine lattice band structure.} 
    (a) Model diagram of the Lieb lattice under a periodic potential. 
    The unit cell vectors of the pristine Lieb lattice, i.e. the lattice in the absence of a periodic potential, are $\bm{a}_1$ and $\bm{a}_2$ where $|\bm{a}_1|=|\bm{a}_2|=a$. 
    The superlattice cell vectors are $\bm{l}_1$ and $\bm{l}_2$, where the superlattice periodicity $L$ is measured in the number of unit cells $N$ such that $L=|\bm{l}_1|=Na$. 
    In the model shown, the periodicity is $L=4a$ and the potential height is $V_0$, while the well and barrier regions are equal in length, $W_w = W_b$.
    Nearest-neighbour $(t)$ and next-nearest neighbour $(t')$ hoppings are shown, in addition to a spin-orbit coupling interaction $(t_{\text{soc}})$; the $\nu_{ij} = \pm 1$ corresponds to an anti-clockwise/clockwise turn between the A and C sites. The shared areas depict regions with $V(x)=V_0$.
    (b) Band structure of the pristine Lieb lattice with nearest-neighbour hoppings only. 
    The high symmetry points of the pristine lattice first Brillouin zone $\bm{\Gamma}(0,0)$, \textbf{X}$(\frac{\pi}{a}, 0)$, \textbf{M}$(\frac{\pi}{a}, \frac{\pi}{a})$, \textbf{Y}$(0, \frac{\pi}{a})$ are shown. The flat band intersects the crossing point of the Dirac-like cone at \textbf{M}.}
    \label{fig:lieb_bands_and_model}
\end{figure}
%
%
Fig.~\ref{fig:lieb_bands_and_model}a shows how we construct the tight-binding (TB) model of the Lieb superlattice (SL). 
In this section we include nearest-neighbour (NN) site interactions only and implement a numerically sharp potential profile. 
The pristine lattice unit cell is spanned by $\bm{a}_1$ and $\bm{a}_2$. In each superlattice unit cell defined by the vectors $\bm{l}_2=\bm{a}_2$ and $\bm{l}_1=N\bm{a}_1$, we set the onsite energies of $L/2$ pristine lattice unit cells  to zero, and the next $L/2$ cells to $V_0$, as shown in Fig.~\ref{fig:lieb_bands_and_model}a. Here, $L=Na$ is the wavelength of the 1D superlattice. We restrict ourselves to the study of symmetric potentials, such that the width of the well and barrier regions are $W_w = W_b = L/2$.
This forms the SL unit cell; we then apply periodic boundary conditions. 
The used Lieb lattice TB model is shown explicitly in Methods.
Fig.~\ref{fig:lieb_bands_and_model}b shows the band structure of the pristine Lieb lattice, which features the intersection of a perfectly flat band with the Dirac-like cone at the high-symmetry \textbf{M} point. 
One requirement of this intersection is the $C_4$ rotational symmetry of the system. 
This is reduced to a $C_2$ symmetry following the application of the periodic potential as outlined in Fig.~\ref{fig:lieb_bands_and_model}a, which destroys this intersection. 
However, it also introduces additional features to the band structure, which we discuss now.
\begin{figure}[t!]
     \centering
     \includegraphics[width=\textwidth]{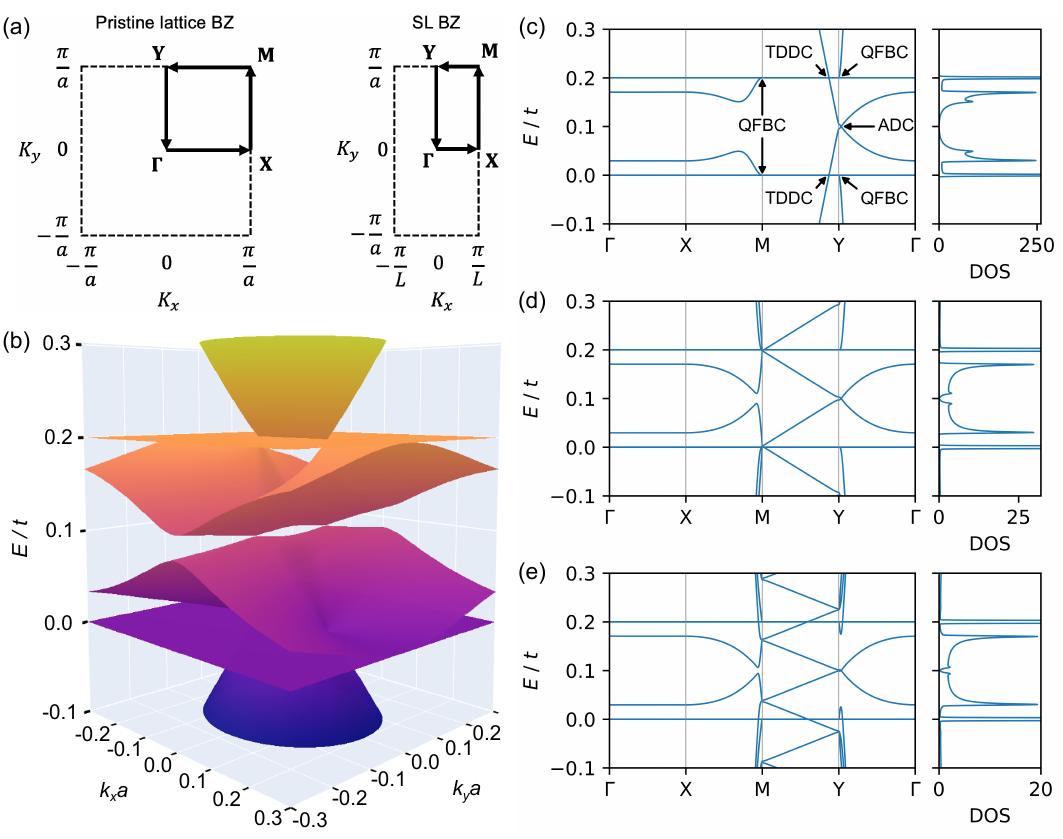}
    \caption{\textbf{Increasing the SL periodicity from the discrete towards the continuum limit.} 
    (a) Pristine and SL first Brillouin zones (BZ) with marked high-symmetry points. 
    (b) The electronic bands for the $L=4a$ superlattice plotted around the \textbf{Y} point. 
    Throughout the paper wave vector components $k_x$ and $k_y$ are relative to the \textbf{Y} point at $(0,\pi/a)$, which remains a high-symmetry point of the SL.
    (c)-(e) Band structures along high-symmetry directions of the SL BZ (left panel) and BZ-integrated density of states (DOS) (right panel) for values of SL periodicity equal to (c) $L=4a$, (d), $L=32a$, (e) $L=50a$. 
    Potential barrier height $V_0 = 0.2 t$. 
    The reciprocal space distances $\bm{\Gamma}$\textbf{X} and \textbf{MY} are artificially kept constant for visualisation of the band-folding along \textbf{MY}. 
    The DOS broadening parameter $\delta = 10{}^{-3} t$.}
    \label{fig:discrete_limit}
\end{figure}
%
%
%

We show in Fig.~\ref{fig:discrete_limit}a the pristine and SL first Brillouin zones (BZs). The \textbf{Y} point located at $(0, \pi/a)$ remains a high-symmetry point, while the \textbf{M} and \textbf{X} points shift due to the SL periodicity. This results in the folding of the pristine bands, which contributes with new states at the \textbf{Y} point. Figure \ref{fig:discrete_limit}b shows the low energy bands of the Lieb SL near the \textbf{Y} point for $L=4a$ and $V_0=0.2 t$. 

In Figs.~\ref{fig:discrete_limit}(c-e) we plot the band structures and BZ-integrated density of states (DOS) for $L=4a$, $32a$ and  $50a$  respectively. 
A continuous evolution of the band structures from $L=2a$ to $L=400a$ is shown in the Supplementary Video (SV) 1 \cite{figshare_doi}. 
There are now two flat bands (FBs) at $E=0$ and $E=V_0$ (i.e. $E/t=0.2$ in Fig.~\ref{fig:discrete_limit}). These arise from the destructive interference of wave functions in the well and barrier regions and manifest as large peaks in the DOS. 
The FBs are degenerate, with the total number, $N_{\text{FB}} = \frac{L}{a}-2$, equal to the number of boundaries between unit cells with the same potential energy (for $L=2a$ there are no perfect flat bands that extend throughout the whole BZ). 
Consequently, the height of the FB-related DOS peaks relative to all other features in the BZ-integrated DOS increases with increasing SL periodicity. 

Breaking the $C_4$ rotational symmetry results in highly anisotropic states that intersect the FBs at $E=0$ and $E=V_0$. 
Marked in Fig.~\ref{fig:discrete_limit}c are triply-degenerate Dirac cones (TDDCs) and quadratic flat band crossings (QFBCs), with the former arising from the folding of the pristine lattice cone along \textbf{YM}. 
The triple-degeneracy becomes apparent in Fig.~\ref{sup_fig:around_TDDC}, where we plot cuts of the band structure around the crossing point for different cut angles $\phi_{\bm{q}}$ (where $|\bm{q}|$ is measured from the crossing and $\phi_{\bm{q}} = 0$ is defined along \textbf{YM}). 
Technically, the degeneracy of the TDDC will increase with $L$, as the number of FBs increases. 
However, the TDDC will be comprised of two dispersing states and the FBs regardless of the value of $L$. 

Additionally, there are two sets of QFBCs. 
The first is at \textbf{Y}, where the original linear crossing point is split, and the maximum (minimum) of the quadratic bands touches the flat bands at $E=0$ ($V_0)$, shown in Fig.~\ref{sup_fig:around_Y}. 
The second set is at \textbf{M}, where the maximum (minimum) cross the FBs at $E=0$ ($V_0$), shown in Fig.~\ref{sup_fig:around_M}. 
It has been shown theoretically that QBCPs can support non-trivial interacting states such as ground-state or nematic ferromagnetism \cite{tsai2015interaction}, owing to the high density of states which is enhanced here due to the crossing with degenerate flat bands. 
In these studies, however, the quadratic band is fully symmetric around the crossing point. 
Here, the bands are only quadratic along $k_y$, potentially opening up the study of directionally dependent interacting phases in the Lieb SL. 
In Figs.~\ref{fig:discrete_limit}(c-e) (and SV1 \cite{figshare_doi}) we show the effect of varying the potential periodicity $L$ on the band structure due to band folding. 
From $L=4a$ to $L=32a$, the bands fold down where an additional two quadratic bands cross the flat bands at \textbf{M}. 
Here, the locations of previously described QFBCs switch between \textbf{M} and \textbf{Y}. 

Finally, the bands that are flat along $\bm{\Gamma}$\textbf{X},  disperse along $\bm{\Gamma}$\textbf{Y} and cross at $\bm{k}=(0, \frac{\pi}{a}-\frac{V_0}{2at})$ and energy $E=V_0/2$, forming an anisotropic Dirac cone, marked as ADC in Fig. \ref{fig:discrete_limit}c.
The band anisotropy around the ADC is shown in Fig. \ref{sup_fig:around_ADC}. 
The ADC is not predicted to exist by previous analytical calculations using a continuum long wavelength model \cite{xu2014omnidirectional}. 
This is because the details of the atomic structure and the left-right discrete lattice symmetry breaking across the well-barrier interface are lost when performing long-wavelength approximation calculations. 
We have found that the ADC crossing is a result of a specific order of coupling between the sites in the SL cell, valid for all values of $L$. 
In the $n^{\text{th}}$ unit cell $(n=1, 2, ..., N)$, the A and B sites couple anti-symmetrically (symmetrically) in the well (barrier) regions, and the C sites then couple anti-symmetrically in consecutive unit cells. 
We give more details on this in the Supplementary Material, and show how an effective description of the bands in the vicinity of the crossing can be derived following a unitary transformation of the original TB Hamiltonian. 
%
%
\section*{The continuum limit}
\begin{figure}[t!]
    \centering
    \includegraphics[width=\textwidth]{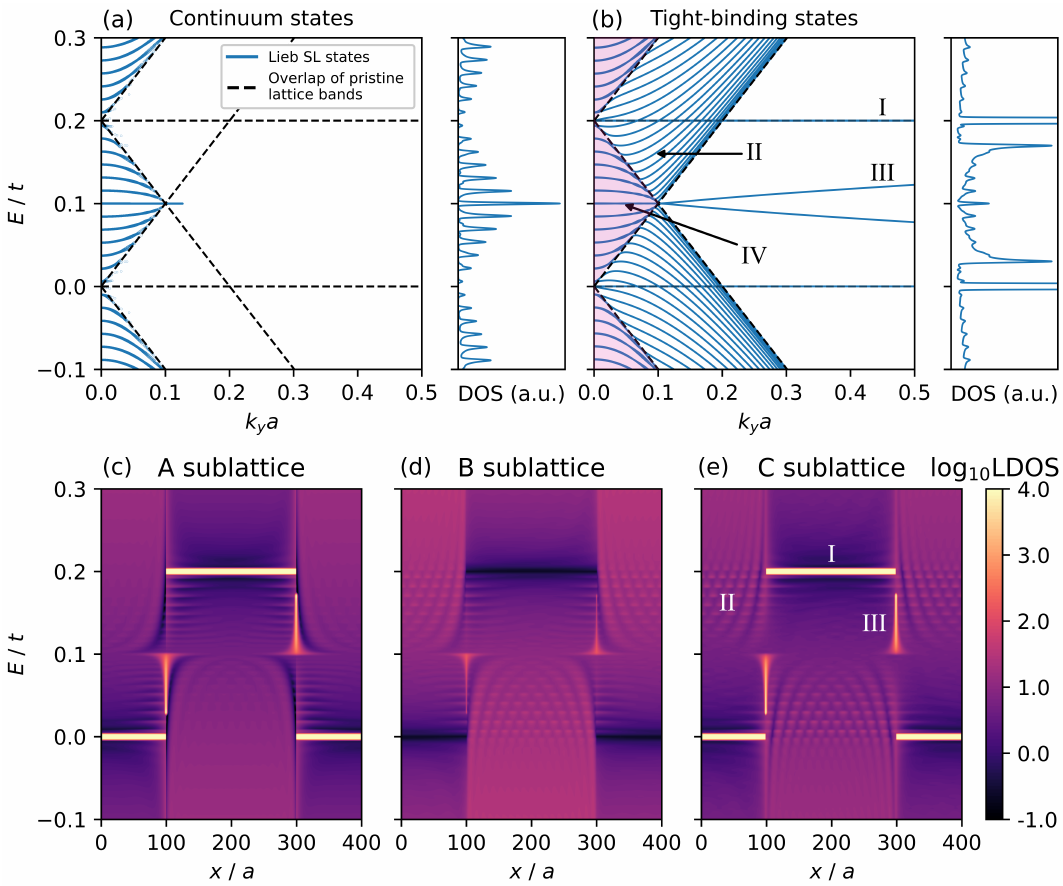}
    \caption{\textbf{Electronic states of the Lieb SL near the }Y\textbf{ point.} 
    The SL parameters are $L=400a$ and $V_0=0.2t$. 
    Band structures along $k_y$ with $k_x=0$ measured from \textbf{Y} in left-hand panels from (a) continuum and (b) tight-binding models with corresponding DOS in right-hand panels. 
    In the band structure plots, the black dashed lines represent the pristine Dirac cones at $E=0$ and $E=V_0$. The overlap of these two pristine lattice dispersions, which bound the continuum states, are shaded in pink. 
    States of interest marked (I-IV) are discussed in the text. 
    The DOS is evaluated by integrating over the states along \textbf{Y}$\bm{\Gamma}$ only.
    Panels (c)-(e) show the real-space LDOS$(x, E)$ along one supercell on the A, B, and C sublattices respectively, to illustrate the localisation of the wave function. LDOS features of states (I-III) affect the LDOS at all sublattices but are more prominent for sublattice C. The LDOS is evaluated by fully integrating over the whole BZ.
    The DOS/LDOS broadening parameter $\delta = 10^{-3}t$.}
    \label{fig:continuum_vs_TB_figure}
\end{figure} 
We now discuss the electronic states of the Lieb SL in the continuum limit, i.e. when $L\gg a$, and show that the full tight-binding (TB) description differs from previous analytical predictions \cite{xu2014omnidirectional}. 
Numerically, we implemented the sharp step potential described in Methods, while the continuum low-energy long-wavelength calculation is done within the transfer matrix (TM) formalism following the methodology presented in Ref.~\cite{cunha2021band} (details in Supplementary Material). 
Note that none of the discrepancies we show are due to numerically resolving scattering between inequivalent valleys (like in the case of graphene) since all high-symmetry points in the Lieb SL BZ can be connected by reciprocal lattice vectors. 
In Fig.~\ref{fig:continuum_vs_TB_figure}(a-b), we compare the analytical and TB band structures for $k_y$ near the \textbf{Y} point and $k_x=0$.
Firstly, we note that the analytical and TB descriptions agree near the \textbf{Y} point where the shifted pristine lattice dispersions overlap (shown by the dashed black lines and the pink shaded regions); here lie extended propagating states. 
In Fig.~\ref{sup_fig:continuum_vs_tb_all_kx}, we show that by superposing the band structures for multiple $k_x$ values as in \cite{xu2014omnidirectional, cunha2021band}, the bands shift in energy and one sees a continuum of states. 
For this reason, we refer to this as the `continuum region'.

\begin{figure}[t!]
    \centering
    \includegraphics[width=\textwidth]{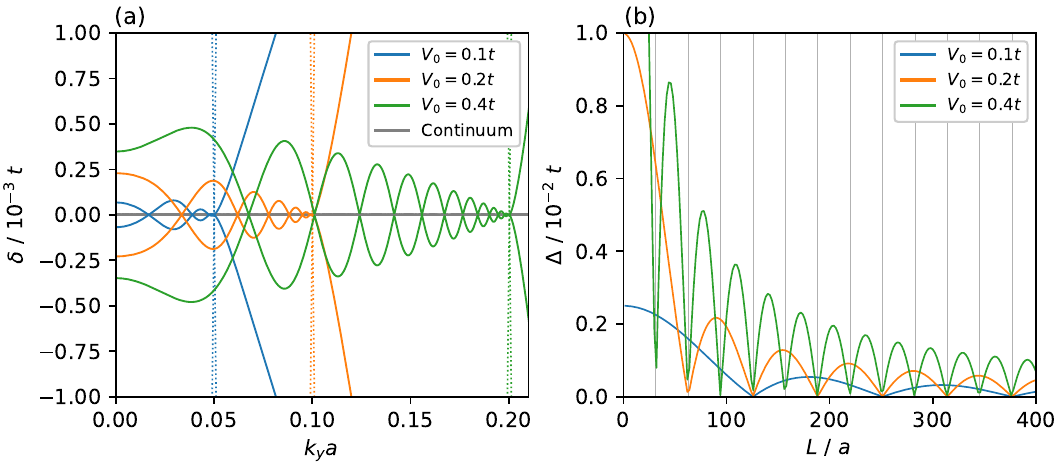}
    \caption{\textbf{The tight-binding approach does not predict the existence of the super-Klein tunnelling (SKT) state.} 
    (a) Flat band expected in the long wavelength model (grey line) and TB energy bands about $E=V_0/2$ for different potential heights $V_0=0.1 t$ (blue line), $V_0=0.2 t$ (orange line), and $V_0=0.4 t$ (green line) with $L=400a$. 
    The dotted lines of the corresponding colour show the overlap of the two pristine lattice cones shifted by that value of $V_0$. 
    The crossing point bounds the validity of the continuum calculation. 
    (b) Gap opening at the \textbf{Y} point of braiding states. 
    The vertical dotted lines show the value of $L$ required to generate an additional crossing for the $V_0=0.4 t$ curve in (a). 
    An additional crossing appears as the gap at \textbf{Y} closes and reopens.}
    \label{fig:skt_states}
\end{figure}
%
There the similarities end. 
The analytical dispersion relation for the SL is the following in the continuum limit\cite{cunha2021band}:
\begin{equation}
    \cos(k_x L) = \cos(k_{x, w}
    W) \cos(k_{x, b} W) - \frac{1}{2} \sin(k_{x, w} W) \sin(k_{x, b}W) \left(\frac{\cos\phi}{\cos\theta} + \frac{\cos\theta}{\cos\phi}\right). \label{eq:continuum_dispersion}
\end{equation}
Here, $k_{x,w}$, $k_{x,b}$ and $\phi$, $\theta$ are the $k_x$ components and propagation direction in the well and barrier respectively. 
Explicit definitions and a derivation of Eq. (\ref{eq:continuum_dispersion}) are given in the Supplementary Material.

The analytical calculation does not contain the flat bands, labelled I in Fig.~\ref{fig:continuum_vs_TB_figure}b. 
The dispersion relation, \ref{eq:continuum_dispersion}, can only be derived from the low-energy conical eigenstates of the pristine lattice band structure. 
If one tries to derive an equivalent dispersion relation using the pristine lattice flat band eigenstates, the inverse of Eq.(\ref{sup_eqn:wronskian_defn}) diverges, as the band index $\alpha=0$. 
This can also be argued physically since the states derived from a TM calculation are those which propagate through the SL, but the flat bands are not propagating states, but are rather localised at the A and C sublattices. 
They are the result of destructive interference of the Bloch states on the B sublattice. 
This is seen in Figs.~\ref{fig:continuum_vs_TB_figure}(c-e), where the large primary peaks in the DOS at $E=0$ and $E=V_0$ due to the localisation of the wave function in the well and barrier regions on the A and C sublattices are missing for the B sublattice. 

The states labelled II in Fig.~\ref{fig:continuum_vs_TB_figure}b, which are quantised states confined to the well/barrier regions, are also missing from the analytical continuum calculation. 
In this region of the band structure, the wavevectors $k_{x,w}$ and $k_{x,b}$ become imaginary, depending on whether states are allowed in the barrier or well at the respective energies in the pristine dispersions. 
The solution of the long wavelength model becomes numerically unstable and diverges, such that resolving these states is computationally unfeasible. 
These quantised states do not shift as $k_x$ is varied, unlike the states in the continuum region indicated with the pink shading in Fig.~\ref{fig:continuum_vs_TB_figure}, meaning they do not propagate along the $\bm{a}_1$ direction. This can be also visualised in Fig. ~\ref{sup_fig:continuum_vs_tb_all_kx}b, where the dispersion is plotted for a range of $k_x$. 

As in Fig.~\ref{fig:discrete_limit}(c-e), the states labelled III are the interface states generated by the discrete lattice symmetry breaking (DLSB) of the system at the well/barrier interface.
The BZ-integrated LDOS shows that these interface states are asymmetric with respect to the SL unit cell, and highly localised to the well-barrier interface, with the left-right localisation asymmetry a consequence of the reduction from $C_4$ to $C_2$ rotational symmetry.
A choice of superlattice cell shifted by $a/2$ flips the energies of these interface states. 
As before, the secondary peaks in the DOS are a signature of these dispersing states and the partial flat bands along $\bm{\Gamma}$\textbf{X}.

A further discrepancy is the predicted nature of the states at $E=V_0/2$, marked IV in Fig. \ref{fig:continuum_vs_TB_figure}b. 
Analytical continuum limit calculations of pseudospin-1 particles tunnelling through a single potential barrier predict $T=1$ transmission for all incidence angles, coined super-Klein tunnelling (SKT).
The band structure of the Lieb SL, i.e. the limit of many barriers, should then feature these states.
For $k_x=0$ and $E=V_0/2$, Eq. (\ref{eq:continuum_dispersion}) becomes $1 = \cos^2(\sqrt{(V_0/2)^2-k_y^2}W) + \sin^2(\sqrt{(V_0/2)^2-k_y^2}W)$, which is satisfied for arbitrary $k_y$. 
This means, as shown in Fig.~\ref{fig:continuum_vs_TB_figure}a, the SKT state manifests as a flat band in the continuum model. 
However, this is not predicted by the TB calculation. 
Plotting states IV for small energies $\delta$ about $E=V_0/2$ in Fig.~\ref{fig:skt_states}a shows that the SKT state is described in fact by two bands which braid to form multiple anisotropic Dirac cones, reminiscent of those found in graphene SLs. 
In fact, the location of these additional crossings follows the same scaling law as found in graphene lattices, $k_y = \sqrt{\frac{V_0^2}{4} - \frac{4n^2\pi^2}{L^2}}$ \cite{barbier2010extra}. 
Therefore, at $E=V_0/2$, we expect the transport properties of the Lieb SL to not resemble SKT, but instead that of graphene under a 1D periodic potential. 
We argue that this is a further consequence of resolving the discrete symmetry broken states (labelled III in Fig. \ref{fig:continuum_vs_TB_figure}b) at the well-barrier interface, and the connecting of the two-band crossings to these states results in a pseudospin-1/2 symmetry, not pseudospin-1. 
Additionally, Fig. \ref{fig:skt_states}b shows the additional crossings are sourced by the gap closing at \textbf{Y}, a result of the band folding along \textbf{YM} that we can resolve using the full TB description of the system. 
Finally, we note that the minima (maxima) of $\Delta$ in Fig, \ref{fig:skt_states}b coincide with the folding of the TDDP onto the \textbf{Y} (\textbf{M}) point shown in Figures \ref{fig:discrete_limit}(c-e) and SV1 \cite{figshare_doi}. 
This in turn coincides with the switching of the curvature of the QFBCs at \textbf{Y} and \textbf{M} discussed in the previous section.
%
%
\section*{Smoothing the potential}
\begin{figure}[H]
    \centering
    \includegraphics[width=\textwidth]{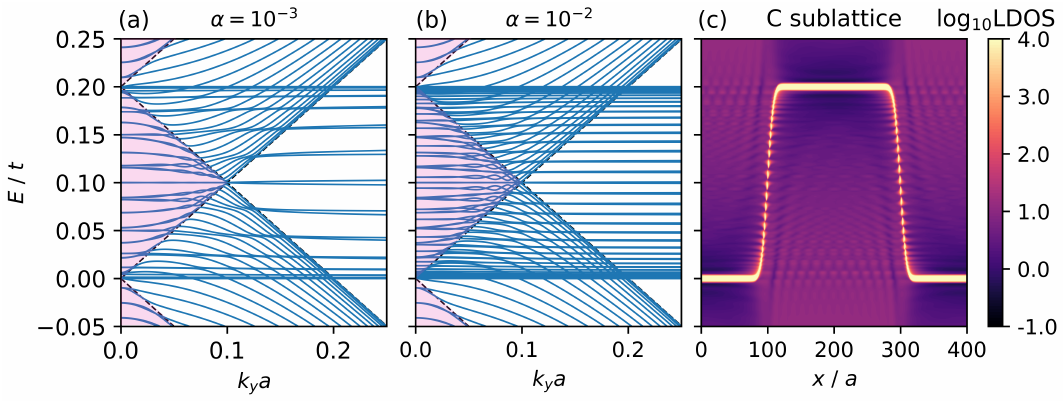}
    \caption{\textbf{Smoothing the potential profile.} The SL parameters are $L=400a$ and $V_0=0.2t$. 
    Low-energy band structures along $k_y$ measured from the \textbf{Y} point and at $k_x=0$, for smoothness parameters (a) $\alpha = 10^{-3}$ and (b) $\alpha = 10^{-2}$. 
   When compared with Figure \ref{fig:continuum_vs_TB_figure}(b) it is clear that smoother potentials lift the degeneracy of the flat bands. As shown in (b) relative to (a), smoother potential steps give rise to more non-degenerate states. 
    The LDOS$(x, E)$ spectra on the C sublattice corresponding to $\alpha = 10^{-2}$ is shown in (c), indicating the localization of these new states and their interaction with the extended and confined states.
    The pink shaded area indicates regions where extended states are allowed in both regions and are defined by the overlap between conical dispersion of the pristine lattice in the well and barrier regions shown by the black dashed lines. }
    \label{fig:smoothed_potential_figure}
\end{figure}
%
The sharp potential profile, i.e. where the potential change occurs over a distance smaller than the lattice constant as depicted in Figure \ref{fig:lieb_bands_and_model}(a), should accurately describe photonic and artificial Lieb SLs \cite{taie2015coherent, vicencio2015observation, huang2011dirac, guzman2014experimental, moitra2013realization}.
However, for those realised using optical potentials or in solid-state systems (through electrostatic gating or dielectric patterning of a substrate), it is unlikely this step change will occur over such a small distance. 
To approximate this, we applied a smoothing of the potential using the sigmoid-like \textit{smoothstep} function to interpolate the potential in each unit cell ${n=(1, \ldots, N)}$ between $V(x, \alpha)=0$ and $V(x, \alpha)=V_0$ for a smoothness parameter value $\alpha$. Details of the exact numerical implementation are given in the Methods section. 
Smoothing the potential profile does not change the $C_2$ rotational symmetry of the system, hence the symmetry-protected crossings (TDDCs, QFBCs and ADC) remain.
However, we show here that it reduces the degeneracy of the flat bands and induces further discrete translational symmetry breaking at the well-barrier interface, giving rise to localised states in the barrier at intermediate energy range $0 < E < V_0$. 

Figs.~\ref{fig:smoothed_potential_figure}(a-b) show the band structures near the \textbf{Y} point for two values of the smoothness parameter $\alpha$, while Fig.~\ref{fig:smoothed_potential_figure}c shows the LDOS spectra across the superlattice unit cell for sublattice type $C$, similar to Fig.~\ref{fig:continuum_vs_TB_figure}e but for a smoothing parameter $\alpha=10^{-2}$. 
The flat band degeneracy lifting shown in Figs.~\ref{fig:smoothed_potential_figure}(a-b) can be seen as a consequence of assigning a different on-site energy to the $n^{\text{th}}$ unit cell over which the potential change occurs. 
While for a sharp potential the degeneracy of a flat band is given by $N_{\text{FB}}=\frac{L}{a}-2$, the smoothing of the potential over several unit cells makes it that a new $V_n(x)$ potential will be assigned to every unit cell $n$, which shifts the on-site energy and reduces the flat band degeneracy and its contribution to the DOS.
However, this feature shows not only as a simple shift of flat bands to new energies, intersecting the extended and evanescent states for each different $V_n(x)$.
Instead, as shown in Fig.~\ref{fig:smoothed_potential_figure}c, multiple tunnelling steps at smaller energy scales appear at the boundary between the $n^{\text{th}}$ and $n+1^{\text{th}}$ cell, generating additional pairs of discrete symmetry broken states similar in nature to states of type III introduced in the previous section. 
These interact and are hybridised with both the evanescent and extended states. 
The dispersion of the former is altered, evidenced by the smearing of the quantised states in the LDOS spectra shown in Fig.~\ref{fig:smoothed_potential_figure}c and in Fig.~\ref{sup_fig:ldos_smoothed_all_sublattices}(a-c) where the LDOS is shown separately for all three sublattices.
Additionally, the dispersion and braiding behaviour of the extended states also changes, shifting the locations of the band crossings, with the effect being more pronounced for smoother SL potentials.

The smoothing of the potential also alters the band structure along $\bm{\Gamma}\textbf{X}$ and \textbf{MY}, as shown in Fig. \ref{sup_fig:smoothed_potential_full_bands}(a-c). 
Each additional pair of discrete symmetry broken state connects to two partial flat bands along $\bm{\Gamma}\textbf{X}$ that reconnect forming doubly degenerate partial flat bands along \textbf{MY}. 
These intersect along \textbf{MY} the linear folded bands, the gradient of which and hence associated group velocity remains unchanged compared to the sharp potential case, forming additional TDDCs between $0 < E < V_0$; smoother potentials generate a greater number of TDDCs. 

Finally, we note further changes to the ``SKT'' state, shown in Fig.~\ref{sup_fig:smoothed_potential_skt_states}(a-c), as compared to the sharp potential case described in Fig.~\ref{fig:skt_states}. 
Smoothing the potential will shift the locations of the ADC towards the \textbf{Y} point: the scaling relation that governs the number and locations of the cones in the sharp potential case is no longer valid and larger values of $\alpha$ suppress the closing of the gap at $E=V_0/2$ that generates them.  
The dispersion around the ADCs is also affected, leading to larger Fermi velocities at larger $k_y$ when compared to the sharp potential case. At the same time the dispersion near the \textbf{Y} point becomes flatter. 
%
%
%
\section*{Adding more parameters}
\begin{figure}[H]
    \centering
    \includegraphics[width=\textwidth]{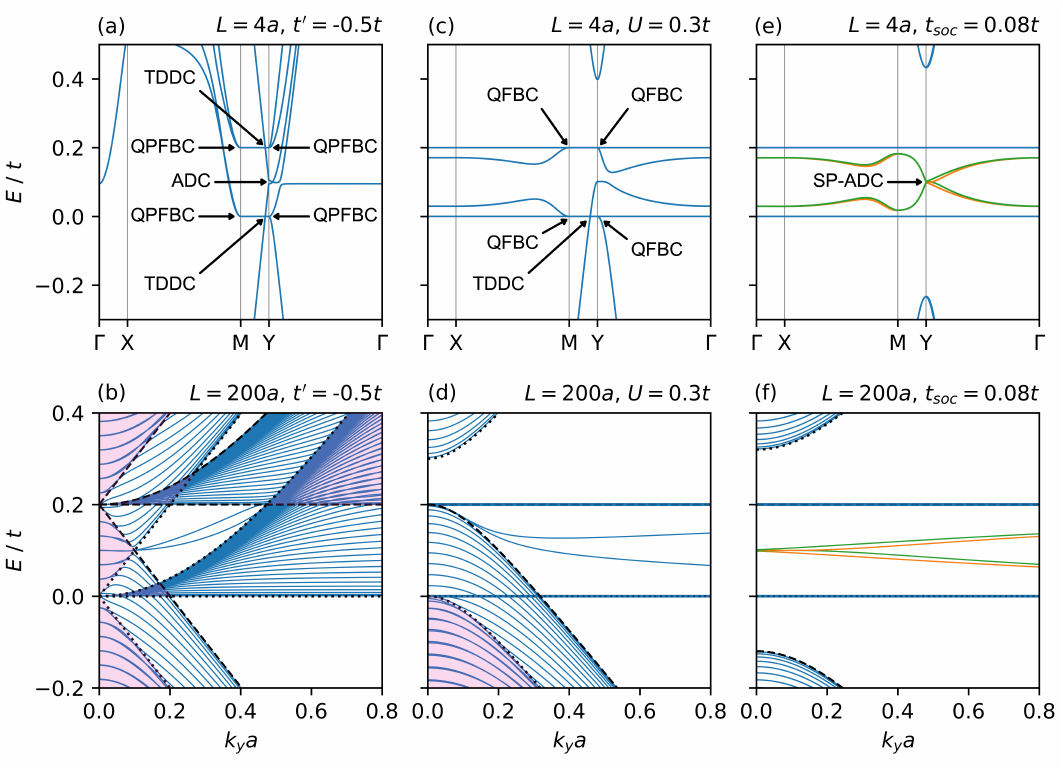}
    \caption{\textbf{Including additional model parameters.} 
    (Top row) Band structure along the high-symmetry path shown in Fig. \ref{fig:discrete_limit}b for $L=4a$ (discrete limit). 
    (Bottom row) Band structure near the \textbf{Y} point for $L=200a$ (continuum limit). 
    The potential height $V_0 = 0.2 t$. 
    From left to right the next-nearest neighbour (NNN) hoppings $t'$, an effective mass term $U$, and an intrinsic spin-orbit coupling (SOC) term $t_{\text{SOC}}$ are individually added to the Lieb SL Hamiltonian.
    The black dotted (dashed) lines represent the pristine lattice dispersions at $k_xa=0$ and $k_xa=\pi$ in the well (barrier) regions, given the corresponding $t'$, $U$ and $t_{SOC}$. The shaded pink regions show where electronic states are allowed in both well and barrier regions according to their respective pristine dispersions. 
    In panels (c) and (f) the green (spin up) and orange (spin down) bands show the spin polarisation of the discrete lattice symmetry broken states in the presence of a SOC term. 
    } 
    \label{fig:adding_single_parameters}
\end{figure}
%

To extend the applicability of our model to solid-state systems, we now include three additional terms in the TB Hamiltonian: (i) a next-nearest neighbour (NNN) interaction term $t'$, to account for non-zero hopping between A/C site orbitals; (ii) an effective mass term $U$, to approximate the chemical potential difference between corner (B) and centre-edge (A/C) sites; and (iii) a spin-orbit coupling (SOC) term $t_{\text{SOC}}$, to extend the model for Lieb lattices hosting metallic elements with high-energy electrons. 
In both Fig.~\ref{fig:adding_single_parameters} and SVs (2-4) \cite{figshare_doi} we show how these parameters individually affect the band structure from the discrete to the continuum limit. 

\textit{Next-nearest neighbour hopping, $t'$}. First, we address the effect of including a NNN hopping interaction ($t'=-0.5t$), likely to be present in all Lieb lattice systems, both artificial and solid state.
As for the pristine lattice case seen in Fig~\ref{sup_fig:pristine_bands_vary_params}a, the presence of the NNN term disperses the original flat bands, which occurs for all wave vectors except along the \textbf{MY} direction. 
Consequently, in the SL there is a significant reduction to the DOS at $E=0$ $(V_0)$, further compounded by the band folding which reduces the distances $\mathbf{\Gamma}$\textbf{X} and \textbf{MY} by a factor of $L/a$. 
However, the flat band degeneracy along \textbf{MY} scales linearly with $L/a$.
As a result, in the discrete limit the contribution to the DOS from the partial flat bands at $E=0$ $(V_0)$ will remain constant for different SL periodicities. 
Furthermore, as seen in Fig. \ref{fig:adding_single_parameters}b and SV2\cite{figshare_doi}, the uneven dispersion of the original degenerate flat bands for $L>4a$ means that in the continuum limit the flat bands at $E=0$ $(V_0)$ are recovered, further enhancing the DOS at these energies. 
%

Even though the original flat band now becomes partially flat in the full BZ, the TDDCs and QFBCs are preserved. We relabel the latter as quadratic \textit{partial} flat band crossings (QPFBCs) owing to the fact that the dispersion of the original flat bands adds further anisotropy to the bands around both the \textbf{M} and \textbf{Y} points, as shown in Figs.~\ref{fig:adding_single_parameters}a and \ref{sup_fig:around_TDDC_NNN}-\ref{sup_fig:around_M_NNN}.
Generally, a larger $t'$ yields greater dispersion and therefore larger modification to the electronic transport when compared to the NN case.  
This is true except at $E=0$ ($V_0$) since the band folding along \textbf{MY} remains unchanged and these energies correspond to the extremal points of the now dispersing flat bands. 
It is important to note that in the discrete limit, only transport experiments that probe around $E=0$ will isolate the dynamics of the charge-carriers around the TDDCs and QFBCs, in addition to the presence of any possible strongly correlated phenomena driven by the partial flat band along \textbf{MY}.
In the continuum limit, shown in Fig.~\ref{fig:adding_single_parameters}b, the picture is more complicated. Compared to the NN case, regions where extended states in the SL are allowed  (depicted in Fig.~\ref{fig:adding_single_parameters}b as pink shaded areas) become more complex due to the unevenly dispersing original flat bands. This can be see in more detail in Figs.~\ref{sup_fig:full_bands_NNN_varyL} and \ref{sup_fig:near_Y_NNN_vary_kx} where the dispersion is shown for different values of L and with for a range of $k_x$.

The discrete symmetry broken states, labeled as type III in  Fig.~\ref{fig:continuum_vs_TB_figure}b, that cross to form the ADC, also disperse strongly upon the addition of the NNN term. 
Consequently, the ADC becomes tilted, as shown in detail in Fig.~\ref{sup_fig:around_ADC_NNN}. 
However, this does not affect the identified scaling relation in the NN case that generates additional cones in the SL since the band folding along \textbf{MY} remains unchanged. 
Instead, all the SL induced ADCs are also tilted and are shifted to energies $E \neq V_0/2$. 
This is shown in Fig.~\ref{sup_fig:SKT_state_NNN} and compared to the ``SKT'' states obtained when only NN hoppings are considered. 
As a result, it should not be expected that the electron transport through the Lieb SL with a non-zero NNN interaction at $E=V_0/2$ would resemble SKT, or that of graphene under a 1D periodic potential. 

\textit{Mass term, $U$.} The inclusion of a positive (negative) mass term $U$ to the pristine Lieb lattice will open a gap above (below) the flat band, while the original linear dispersion of the bands becomes quadratic, see Fig~\ref{sup_fig:pristine_bands_vary_params}b. 
For the Lieb SL, the value of $U$ needed to open a gap depends on the SL periodicity: in the discrete and continuum limits these conditions are $|U|>V_0/2$ and $|U|>V_0$ respectively, shown in Fig. \ref{sup_fig:U_limits}. 
A positive (negative) $U$ satisfying these conditions will open a gap above (below) the top (bottom) flat bands, preserving the set of TDDCs and QFBCs at $E=0$ $(V_0)$, shown for $U=0.3t$ in the discrete limit in Fig. \ref{fig:adding_single_parameters}c. 
Additionally, any $|U|>0$ destroys the ADC, as the broken sublattice symmetry prevents the Bloch states on the A and B sublattice combining within the coupling conditions required to generate it. 
%

%
The additional ADCs generated in the continuum limit are also affected. 
Whilst their location in energy $E = (V_0 + U)/2$ agrees with the analytically predicted SKT state for massive charge-carriers \cite{betancur-ocampo_super-klein_2017}, Fig. \ref{sup_fig:SKT_U} shows fewer ADCs as the NN only case for the same SL periodicity and height. 
This is because for finite $U$, the width (in $k_y$) of the overlapping pristine conduction/valence bands from the well/barrier regions shrinks, reducing the number of available wave vectors that can host them. 
Also, the mass term increases the size of the gap at \textbf{Y} which must close to generate additional ADCs, thus a larger $L$ is required to do so for a given $V_0$. 
This deviation from the SKT prediction is greater in this case than when only NN hoppings are included: we do not expect SKT of massive particles in a Lieb SL. 

As for previous cases, the band folding along \textbf{MY} is affected by the inclusion of the mass term as the SL periodicity is increased towards the continuum limit: the gaps between consecutive states fold towards $E=0$ and are flattened (Fig.~\ref{sup_fig:full_bands_U_varyL} and SV3\cite{figshare_doi}).
The slope of these flattened bands is also correlated to the dispersion in the $k_x$  direction, as shown in Fig.~\ref{sup_fig:near_Y_U_vary_kx} where the dispersion is plotted for a range of $k_x$. 
This effectively represents the probability of charge carriers tunnelling through the SL. 
Therefore, if these states do not disperse in the $k_x$ direction, e.g. at $E=0.05t$ for $L=50a$, or $0<E<V_0$ for $L=200a$, this will give rise to super-collimation along $|\bm{a}_2|$ within the barrier. Effectively, electrons are allowed to move only in the $y$ direction. Such configurations can be readily obtained in the limit $V_0<2\Delta$, where $\Delta$ is the mass term induced gap of the pristine Lieb lattice.

\textit{Spin-orbit coupling term}, $t_{SOC}$. In the case of the pristine Lieb lattice, it is well known that the application of an SOC term opens topological gaps (with Chern number $C=\pm1$) above and below the flat bands \cite{springer_topological_2020, goldman_topological_2011}. This can be also seen in the pristine Lieb lattice dispersion in the presence of SOC, shown in Fig.~\ref{sup_fig:pristine_bands_vary_params}c.
In the case of the SL, this modification destroys the TDDCs and QFBCs, as shown in Fig. \ref{fig:adding_single_parameters}(e). 
However, the presence of discrete lattice symmetry breaking in the SL induces spin-polarisation along \textbf{XM} and \textbf{Y}$\mathbf{\Gamma}$ which splits the ADC into two, forming a spin-polarised ADC (SP-ADC).
This is shown for the green (spin-up) and orange (spin-down) spin-split interface bands in Fig.~\ref{fig:adding_single_parameters}e and in more detail in Figs.~\ref{sup_fig:full_bands_SOC_varyL} and \ref{sup_fig:SKT_state_SOC}, where the splitting becomes visible. 
As the SL periodicity is increased towards the continuum limit the two quadratic higher-energy bands shift towards the flat bands, eventually converging to the pristine lattice dispersions obtained for $t_{SOC}=0.08t$, shown by the black dashed (well) and dotted (barrier) lines in Fig.~\ref{fig:adding_single_parameters}f. 
However, unlike the NN and NNN case, increasing the SL periodicity towards the continuum limit does not generate additional ADCs since there are no available extended states to fold at $E=V_0/2$. 
In fact, the SP-ADCs remain stationary as $L$ is varied, as shown in Fig.~\ref{sup_fig:SKT_state_SOC}.
\begin{figure}[t!]
    \centering
    \includegraphics[width=\textwidth]{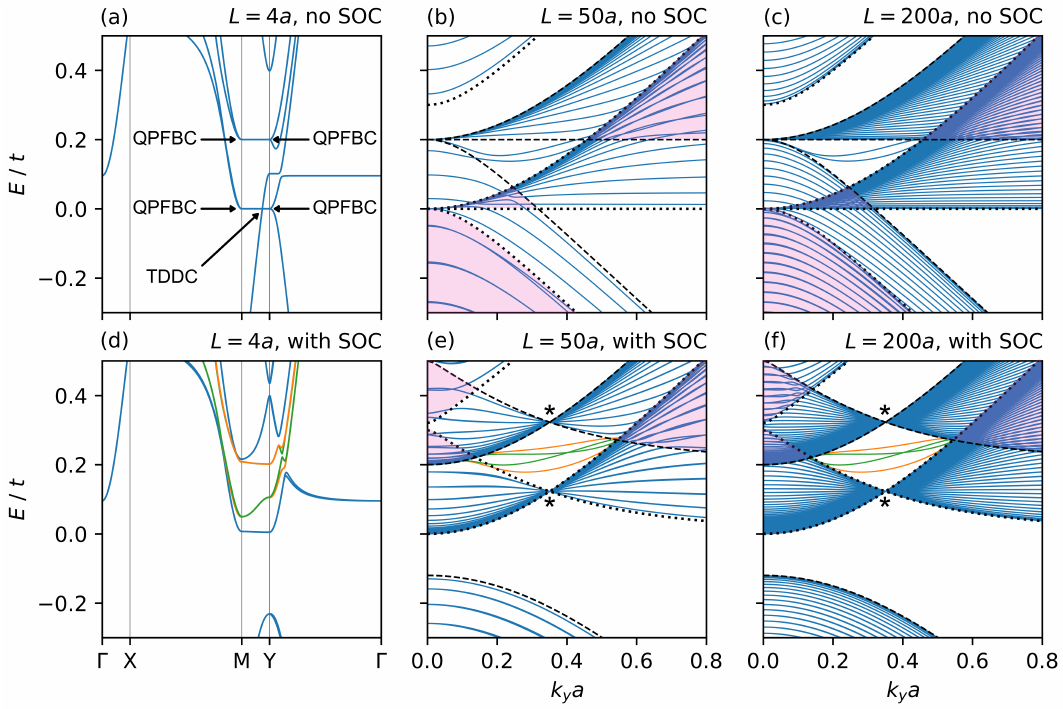}
    \caption{\textbf{Describing realistic solid-state Lieb SLs without and with spin-orbit coupling.} 
    Top row shows (a) the full band structure for $L=4a$ (discrete limit) and (b-c) the band structure near the \textbf{Y} point for $L=50a$ and $L=200a$ and a solid-state system without SOC. 
    Bottom row shows the band structure for the same SL wavelength but including a SOC term. 
    The parameters are $t'=-0.5 t$, $U=0.3 t$, and $t_{\text{SOC}}=0.08 t$.
    In panels (e) and (f) the green (spin up) and orange (spin down) bands show the spin polarisation of the discrete lattice symmetry broken states.
    Same as Fig.~\ref{fig:adding_single_parameters},  the black dotted (dashed) lines are the pristine lattice dispersions at $k_xa=0$ and $k_xa=\pi$ in the well (barrier) regions given the corresponding $t'$, $U$ and $t_{SOC}$ parameters. The shaded pink regions show where electronic states are allowed in both well and barrier regions according to their respective pristine dispersions. 
    }
    \label{fig:adding_all_parameters}
\end{figure}
%

\textit{All terms included}. A realistic and experimentally relevant description of a solid-state Lieb SL system requires at minimum an NNN hopping interaction $t'$ and an effective mass term $U$, simultaneously included.  
The inclusion of an SOC term $t_{\text{SOC}}$ is also required if metallic elements are present. The resulting band structure in the pristine lattice is shown in Fig.~\ref{sup_fig:pristine_bands_vary_params}d.
In Fig. \ref{fig:adding_all_parameters} we show the full band structures for $L=(4, 50, 200)a$ and the states near the \textbf{Y} point for $t'=-0.5t$, $U=0.3t$ without and with an SOC interaction $t_{SOC}=0.08t$.
We also provide a data file containing full band structures for a range of SL parameters which can be visualised using the accompanying Python script \cite{figshare_doi}. 
We observe that in the absence of SOC (Fig. \ref{fig:adding_all_parameters}a), the band dispersions inherit features from configurations with NNN or mass term considered separately, i.e. original flat bands of the Lieb lattice become dispersive except along \textbf{MY} (NNN), the original linear bands become quadratic and gaps are introduced in the spectrum (mass term).  As such, we see the appearance of four QPFBCs and a bottom (top) TDDC for a positive (negative) mass term, together with the disappearance of the ADC at $E=V_0/2$. 

A non-zero SOC interaction will gap these crossings, disperse the partial flat bands along \textbf{MY}, and induce a small spin-splitting of the bands along \textbf{XM} and \textbf{Y}$\mathbf{\Gamma}$, most apparent in the well-barrier interface bands in Fig. \ref{fig:adding_all_parameters}d, shown as orange/green bands for spin-down/up components. 
However, given that these spin-split bands are dispersing and overlap with other non-spin polarized bands, these cannot be individually detected in spin-sensitive transport experiments. 

The SOC interaction can also open a band gap, the nature of which is set by the sign of the mass term: the system is a topological (trivial) insulator for a positive (negative) mass term. 
The topological gap hosts a quantum spin-Hall edge-state corresponding to spin-Chern numbers $C_{\uparrow \downarrow}=\pm 1$.

As the SL periodicity is increased towards the continuum limit, shown in \ref{fig:adding_all_parameters}(b-c) and (e-f), the band structures for either cases without and with SOC become increasingly complex, owing to the existence of continua of extended propagating states (shaded pink areas), the well/barrier localised states (type II),  the interface states (type III) and the uneven dispersion of original degenerate flat bands induced by the NNN term. 
However, there are general features, including band-flattening and the isolation of localised states that arise for broad ranges of Lieb SL parameters in the continuum limit, which we point out here. 
Without SOC, flat bands that span the whole BZ at $E=0$ $(V_0)$ are recovered in the continuum limit (Figs.~\ref{fig:adding_all_parameters}c and \ref{sup_fig:full_bands_NNN_U_varyL}) and the presence of the TDDC at $E=0$ is tuneable with the SL periodicity (see also SV5)\cite{figshare_doi}. 
With the inclusion of the SOC term, this band flattening at $E=0$ is not perfect, but the presence of the topological edge state is expected to impact the nature of many-body correlations present in the nearly-flat bands. 
We also note that increasing the SL periodicity has no effect on the number of topological edge-states. 
This has been verified using ribbon band structure calculations but can also be argued simply: the topological gap does not close and reopen as the SL periodicity is varied, hence the number of edge-states remains unchanged ($C_{\uparrow \downarrow}=\pm 1$). 

SV6\cite{figshare_doi} shows that the spin-up interface states (green bands in Fig. \ref{fig:adding_all_parameters}(e-f)) form a tilted cone at approximately $L=38a$ for the given SL parameters. 
But, for larger SL periodicities, spin-sensitive transport experiments will show even smaller discrepancies than the discrete limit case given the greater number of dispersive states that can contribute to transport.
Figures \ref{fig:adding_all_parameters}(e-f) and SV6 \cite{figshare_doi} also show the formation of two highly-degenerate crossing points (marked *), with the bottom (top) crossing formed from converging well (barrier) localised states. 
The degeneracy of these crossings is $\frac{L}{a} - 3$ and is formed due to the folding of the band structure through a constant energy contour of the pristine lattice band structure. 
Plotting a cut of the band structure along $k_x$ through this crossing shows no dispersion. 
This is another feature where dispersive and flat bands occur at the same energies in the band structure. 
Importantly, these crossings are a result of band folding, a universal feature of all periodic potentials, not a sensitive symmetry constraint of the system.

Finally we note general features relevant to all Lieb SLs in the continuum limit, without and with SOC.
We expect super-collimation of charge carriers along the $|\bm{a}_2|$ direction to be measurable in transport experiments due to both the isolation of the localised states from the continua and the corresponding band flatness along \textbf{MY}. 
This would occur, for example, at energies $E=0.1t$ in Figures \ref{fig:adding_all_parameters}(c, f). 
Additionally, for systems with a negative mass term, insulating band gaps can be opened below $E=0$ permitting switchable on/off transport measurements. 
The near-flatness of the dispersing flat bands shown in Figs.~\ref{sup_fig:full_bands_NNN_U_varyL}-\ref{sup_fig:near_Y_NNN_U_SOC_vary_kx} will enhance the DOS as energies $E>0$ for these super-collimated states and those within the continuum (shaded pink regions), possibly leading to the emergence of strongly correlated phases.  
%
%
\section*{Discussion}
In the following we summarize the results presented in the previous sections. These can be categorized broadly according to the length scale of the SL, e.g. discrete versus continuum limit, or to the addition of relevant new terms to the Lieb lattice Hamiltonian, e.g. next-nearest neighbour hopping term, imbalance in the on-site potential – mass term, or the presence of SOC coupling interactions.

In the discrete limit, i.e. $L \sim a$, the electronic dispersion of the Lieb SL presents a range of unexpected features, such as tilted Dirac cones, intersections between quadratic and (partial) flat bands, or spin-polarized states in the presence of SOC. 
On the other hand, by exploring the continuum limit, i.e. $L \gg a$, we compare with previous simulations performed within the continuum limit at low energies and long wavelengths near the Dirac cone. Several features compare well with previous calculations, e.g. the nature of the extended states in the SL and the appearance of confined states that are allowed only in well or barrier regions, depending on the energy. 
At the same time, we observe new phenomena due to the discreetness of the Lieb lattice and the symmetry breaking induced by the SL. For example, localised states at the well/barrier interface give rise to a series of anisotropic Dirac cones at $E=V_0/2$, the number of which depend on the $L/a$ ratio, reminiscent of periodicity-induced extra Dirac cones in graphene. These states correspond to the SKT states in the continuum approach, but the asymptotic limit $L/a \rightarrow \infty$ is non-trivially approached due to the existence of Dirac cones, albeit with increasing frequency and decreasing Fermi velocity or flattening.
Furthermore, we show that in the realistic case, where the well/barrier interface becomes smooth, additional flat interface states appear in the energy range $0<E<V_0/2$ and eventually hybridize with both the extended and localized states. The interface smoothing also affects the SKT states and ADCs.

Adding additional terms to the Hamiltonian strongly affects the electronic states in the SL. 
For example, the next-nearest neighbour hopping disperses the original flat bands in the BZ, except along the \textbf{MY} direction. This gives rise to additional regions in the BZ where extended states are allowed and induces a larger number of localized states in the well/barrier regions. Furthermore, the SKT states around $E=V_0/2$ become dispersive, resulting in further deviation from the original SKT states.

A mass term gives rise to gapped regions in the spectrum and completely erases the SKT states when $V_0$ is smaller than the pristine gap. We again find regions in the BZ where extended states are possible. Additionally, we find cases showing only localized states within a certain energy range, when no states are allowed in one of the well or barrier regions. Since these do not disperse in the $k_x$ direction, they would give rise to super-collimation resulting in pure propagation only along the $k_y$ direction.

The SOC term gives rise to topological gaps that survive in the SL, and we find spin-polarized interface states that propagate only along the $k_y$ direction.

When all the different terms are combined the band structure becomes complex with shared features resulting from the individual terms: original flat bands become dispersive, (topological) gaps can appear, regions where extended states are allowed shift around the BZ and in energy, well/barrier localized states are ubiquitous and can be isolated within some energy ranges and spin-polarized states can be localized at the well/barrier interfaces.

In conclusion, our results reveal a complex picture regarding the effect of 1D periodic potentials on the electronic properties of Lieb lattices. By considering a TB approach we were able to smoothly transition from the discrete to the continuum regimes, with relevance to possible experimental realizations of the SL periodicities. Furthermore, the TB approach allowed us to include terms in the Hamiltonian that describe experimentally relevant effects. These are related to longer range hoppings, distortions or imbalances in the sublattices or the presence of spin-orbit interactions. These terms give rise to rich features observed in the band structure, relevant to future experiments in artificial Lieb lattices or theoretically predicted COFs.
%
%

\section*{Methods}
\subsection*{Tight-binding Hamiltonian}
Using the Python-based Pybinding package \cite{dean_moldovan_2020_4010216}, we build a tight binding model with a single orbital per site to study the electronic states of a single-layer Lieb lattice under an electrostatic potential applied along the $\bm{a}_1$ direction forming a superlattice (see Fig. \ref{fig:lieb_bands_and_model}a). The Hamiltonian of the system is $H = H_0 + V(x)$, where the pristine lattice Hamiltonian $H_0$ is
\begin{align}
    H_0 = \sum_{i} \epsilon_i c_{i}^\dagger c_{i} 
    + \sum_{i,j} t_{ij} c_i^\dagger c_j 
    + i t_{\text{SOC}} \sum_{i,j} \nu_{ij} c_i^\dagger \sigma^z c_j + H.C. 
\end{align}
where /textit{H.C.} includes the Hermitian conjugate terms. $\epsilon_i$ sets the on-site energy of the A, B, and C sublattices. Here, $\epsilon_A = \epsilon_C$ and $\epsilon_B=U$ is used to approximate the differences in the chemical potentials of the corner and centre-edge sites. The $t_{ij}$ hoppings in the second term are considered up to next-nearest neighbours. The parameter $t_{\text{SOC}}$ sets the strength of the third spin-orbit coupling term and $\nu_{ij} = \pm 1$ corresponds to an anti-clockwise / clockwise hopping between the A and C sublattice sites. $V(x)$ is the electrostatic potential modeled as a spatially varying onsite potential defining the SL well-barrier profile shown in Fig.~\ref{fig:lieb_bands_and_model}a. Then, the momentum space Hamiltonian is constructed in the basis of Bloch wave functions $\psi_{\bm{k}}$ as $H_{\bm{k}} = \psi_{\bm{k}}^\dagger H_0 \psi_{\bm{k}}$ and fully-diagonalised. The resulting band structure for the pristine system can be visualised in Fig.~\ref{fig:lieb_bands_and_model}b and Fig.~\ref{sup_fig:pristine_bands_vary_params}(a-d).
\subsection*{Numerical implementation of sharp and smooth potentials}
We investigate two forms of periodic potential: an atomically sharp step-like potential where the potential change occurs instantaneously, and a potential with a finite smoothness. These are referred to as `step' and `smooth' periodic potentials in the main text. Application of the step periodic potential $V_{\text{step}}(x)$ is achieved using the Heaviside step function $\Theta(x)$ such that 
\begin{align}
    V_{\text{step}}(x) = V_0 \Theta(x) = \begin{cases}
            0, & \quad 0 < x < L/2 \\
            V_0, & \quad  L/2 \leq x \leq  L
        \end{cases}
\end{align}
where the periodicity of the superlattice $L=Na$ is in units of the original lattice unit cell length. The smooth potential is implemented using the sigmoid-like \textit{smoothstep} function $S(x)$. It is a function that takes two end points as its argument and smoothly interpolates using a Hermite polynomial of degree $n$; this tunes how quickly the potential changes through the superlattice. Fixing the end points at $x=0$ and $x=L/2$ the smoothed potential profile $V_{\text{smooth}}(x, \alpha)$ with smoothness $\alpha$ is
\begin{align}
    V_{\text{smooth}}(x, \alpha) = V_0 S_\alpha(x) = \begin{cases}
        0, & \quad x = 0 \\
        V_0 \left(\frac{1}{2} R_\alpha(x) (2x-1) + \frac{1}{2}\right) & \quad 0 < x < L/2 \\
        V_0, & \quad x = L/2
    \end{cases}, 
\end{align}
where $R_\alpha(x)$ is an odd-symmetry polynomial given by 
\begin{align}
    R_\alpha(x) = \left(\int_0^1 (1-u^2)^{\alpha^{-1}}\right)^{-1} \int_0^x (1-u^2)^{\alpha^{-1}} du. 
\end{align}
The odd-symmetry of $R_\alpha(x)$ is then used to build the full potential between $x=L/2$ and $x=L$. 
\section*{Acknowledgements}
Dylan Jones was supported by a Mirai PhD Scholarship in Quantum Phenomena in 2D Materials through the Advancement Office at the University of Bath.

%
\printbibliography[heading=bibintoc, title={References}] 
%
%
\setcounter{figure}{0} 
\renewcommand{\thefigure}{S\arabic{figure}} 
\setcounter{section}{0} 
\setcounter{equation}{0} 
\renewcommand{\theequation}{S.\arabic{equation}} 
\renewcommand{\thesection}{S\arabic{section}} 
 \newpage
\section*{Supplementary Material}

\subsection*{Projection onto the anisotropic Dirac cone (ADC)}
To project the original Lieb SL TB Hamiltonian $H$ onto the ADC crossing point, we need the unitary transformation $\Tilde{H} = UHU^\dagger$ that gives
\begin{equation}
    \Tilde{H} = \begin{bmatrix}
        H_0 & T \\
        T^\dagger & H_1
    \end{bmatrix},
\end{equation}
where $H_0$ is a $2\times 2$ zero-eigenvalue block at the ADC crossing. To find the unitary matrix $U$, the wave function is expressed in a new basis where the first two components are a set, prescribed linear combination of the basis states. 
First labelling the sites in the Lieb SL according to their type (A, B, or C) and what unit cell they sit in $(n=1, 2, ..., N)$, this prescription is as follows:
\begin{enumerate}
    \item take antisymmetric (symmetric) combinations of the $\ket{A_n}$ and $\ket{B_n}$ basis states in the well (barrier) regions, then combine these symmetrically;
    \item antisymmetrically combine every consecutive $\ket{C_n}$ site state throughout the superlattice;
    \item the first (second) components in the new basis are then antisymmetric (symmetric) combinations of the two above results;
    \item the remaining states are chosen to be other, orthogonal combinations of the basis states that numerically evaluate to (approximately) zero to maintain orthonormality.
\end{enumerate}

To demonstrate this we choose the simplest case of $L=2a$. 
Following the procedure outlined above, the wave function written in the new basis is
\begin{equation}
    \ket*{\Tilde{\psi}} = U\ket{\psi} = \begin{bmatrix}
        \left(\ket{A_1} - \ket{B_1} + \ket{A_2} + \ket{B_2}\right) / 2\sqrt{2} + \left(\ket{C_1} - \ket{C_2}\right) / 2 \\
        \left(\ket{A_1} - \ket{B_1} + \ket{A_2} + \ket{B_2}\right) / 2\sqrt{2} - \left(\ket{C_1} - \ket{C_2}\right) / 2 \\
        (\ket{A_1} - \ket{B_1} - \ket{A_2} - \ket{B_2}) / 2 \\
        (\ket{A_1} + \ket{B_1}) / \sqrt{2} \\
        (\ket{A_2} - \ket{B_2}) / \sqrt{2} \\
        (\ket{C_1} + \ket{C_1}) / \sqrt{2}
    \end{bmatrix}. 
\end{equation}
Then, performing the corresponding unitary transformation on the original $L=2a$ Lieb SL Hamtiltonian $H$ gives the $2\times 2$ block 
\begin{equation}
    H_0 = \begin{bmatrix}
        0 & i\sqrt{2} \sin(k_x a/2) \\
        -i\sqrt{2} \sin(k_x a/2) & 0
    \end{bmatrix}, 
\end{equation}
which has the required two zero-eigenvalues at the location of the ADC crossing. 

\subsection*{Continuum model calculation}
The low energy and long wavelength continuum model calculation of the electronic states of the Lieb lattice under a 1D periodic can be done using the Transfer Matrix (TM) formalism. 
The eigenstates of the pristine system at low energies near the \textbf{M} point (see Fig. \ref{fig:lieb_bands_and_model}(b)) are 
\begin{equation}
    \psi_{\alpha}(\bm{r}, \bm{k}) = \begin{bmatrix}
        \sin \phi_{\bm{k}} \\
        \alpha \\ 
        \cos \phi_{\bm{k}}
    \end{bmatrix} e^{i \bm{k} \cdot \bm{r}}, 
\end{equation}
where $\alpha$ is the band index ($+1$ for the conduction band, $0$ for the flat band, $-1$ for valence band), $\bm{k}$ is measured from M, and the propagation direction $\phi_{\bm{k}}=\tan^{-1}(k_y / k_x)$.
Translational invariance along $y$ means the problem reduces to 1D. 
These are matched at the well-barrier interface using the matching conditions 
\begin{equation} 
    \psi_B^+(x) = \psi_B^-(x), \qquad \psi_C^+(x) = \psi_C^-(x), \label{Seq: matching conditions} 
\end{equation}
in order to build the TM $T$. 
The $+$ ($-$) correspond to states approaching the well-barrier interface from the left (right). 
The TM is built using a product of the Wronskians $\mathcal{W}$ of the matched eigenstates at the well-barrier interfaces, given by
\begin{equation}
    T = \mathcal{W}_{k_{x, b}}(L) \mathcal{W}_{k_{x, b}}^{-1}(W) \mathcal{W}_{k_{x, w}}(W) \mathcal{W}_{k_{x, w}}^{-1}(0). 
\end{equation}
Here, $k_{x, w} = \sqrt{\left(\frac{E}{\hbar v_F}\right)^2 - k_y^2}$ and $k_{x, b} = \sqrt{\left(\frac{E-V_0}{\hbar v_F}\right)^2 - k_y^2}$ are the $x$-components of $\bm{k}$ in the well and barrier regions. $\hbar v_F$ is the Fermi velocity of the Dirac bands.
Explicitly, at a position $x$, we have
\begin{equation}\label{sup_eqn:wronskian_defn}
    \mathcal{W}_{k}(x) = \begin{bmatrix}
        \alpha e^{i k x} & \alpha e^{-i k x} \\
        \cos (\phi_{\bm{k}}) e^{i k x} & -\cos (\phi_{\bm{k}}) e^{-ikx}
    \end{bmatrix}. 
\end{equation}
The fact that $\text{Tr}(T) = 2 \cos(k_x L)$, with $k_x$ the superlattice wave vector along $x$, gives the dispersion relation
\begin{equation}\label{sup_eqn:analytical_dispersion_relation}
    \cos(k_x L) = \cos(k_{x, w}
    W) \cos(k_{x, b} W) - \frac{1}{2} \sin(k_{x, w} W) \sin(k_{x, b}W) \left(\frac{\cos\phi}{\cos\theta} + \frac{\cos\theta}{\cos\phi}\right), 
\end{equation}
where $\phi$ and $\theta$ are the propagation directions in the well and barrier respectively.
\newpage
\subsection*{Supplementary figures}
\begin{figure}[H]
     \centering     \includegraphics[width=0.95\textwidth]{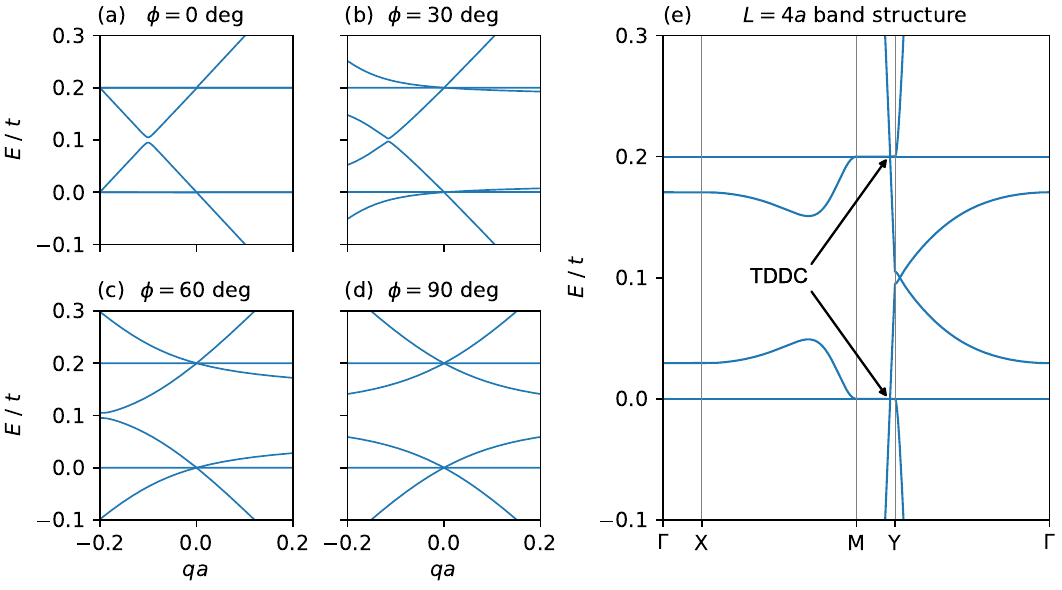}
    \caption{\textbf{Band structure around the triply degenerate Dirac cone (TDDC).} 
    (a-d) The band structures along small reciprocal space distances $q$ measured from the TDDC. 
    In each subplot the angle of this cut is rotated by $\phi$ ($\phi=0$ defined as along \textbf{Y}$\rightarrow$\textbf{M}).
    (e) The low-energy $L=4a$ Lieb SL band structure with TDDCs indicated. 
    }
    \label{sup_fig:around_TDDC}
\end{figure}

\begin{figure}[H]
     \centering     \includegraphics[width=0.95\textwidth]{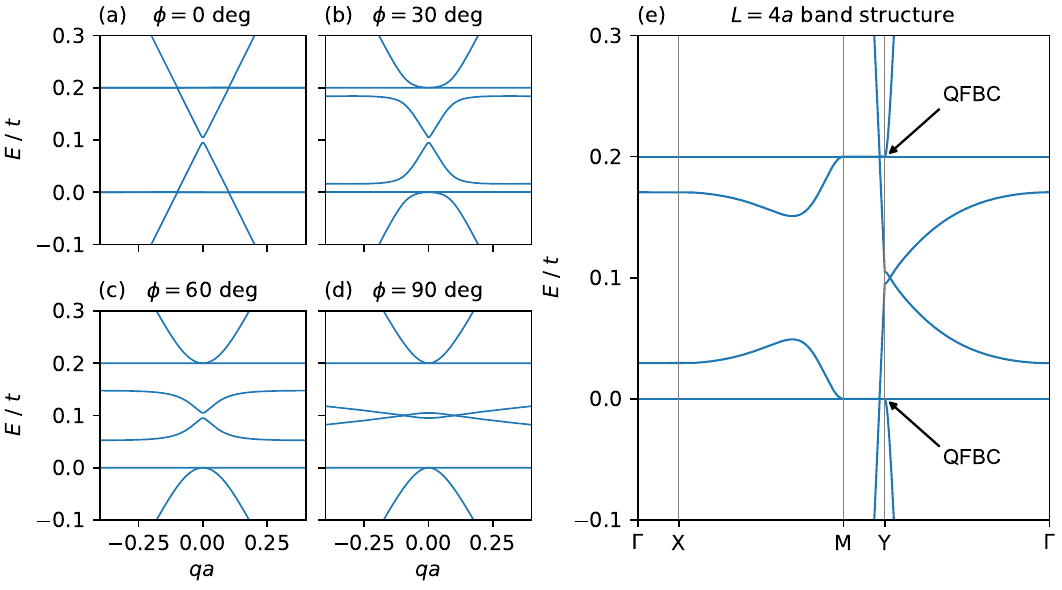}
    \caption{\textbf{Band structure around the Y point.} 
    (a-d) The band structures along small reciprocal space distances $q$ measured from the \textbf{Y} point at $\bm{k} = (0, \pi)$. 
    In each subplot the angle of this cut is rotated by $\phi$ ($\phi=0$ defined as along \textbf{Y}$\rightarrow$\textbf{M}).
    (e) The low-energy $L=4a$ Lieb SL band structure with QFBCs indicated.
    }
    \label{sup_fig:around_Y}
\end{figure}

\begin{figure}[H]
     \centering     \includegraphics[width=0.95\textwidth]{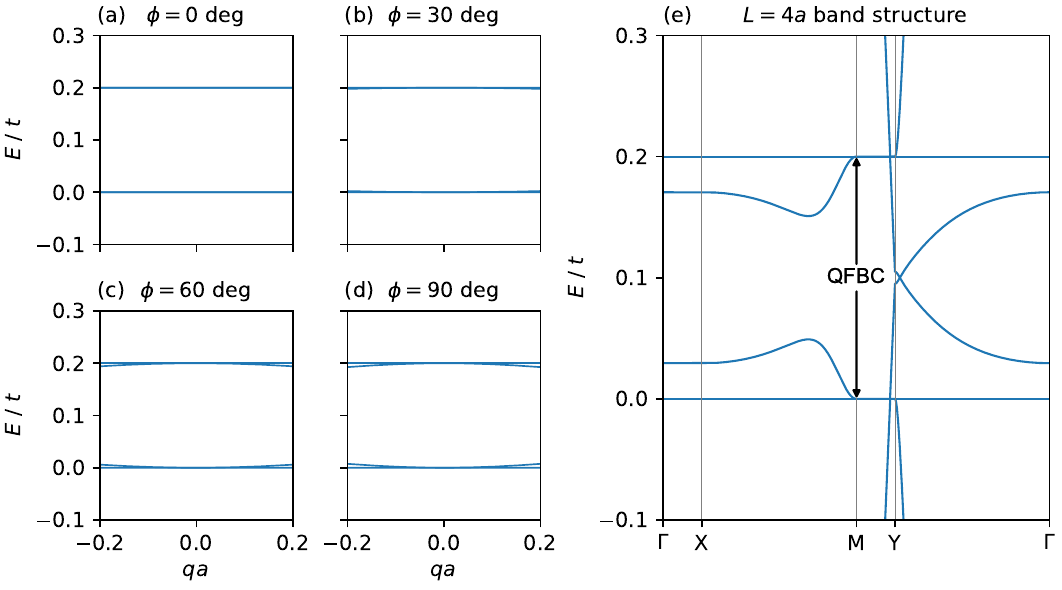}
    \caption{\textbf{Band structure around the M point.} 
    (a-d) The band structures along small reciprocal space distances $q$ measured from the \textbf{M} point at $\bm{k} = (\pi/L, \pi)$. 
    In each subplot the angle of this cut is rotated by $\phi$ ($\phi=0$ defined as along \textbf{Y}$\rightarrow$\textbf{M}).
    (e) The low-energy $L=4a$ Lieb SL band structure with QFBCs indicated.
    }
    \label{sup_fig:around_M}
\end{figure}

\begin{figure}[H]
     \centering     \includegraphics[width=0.95\textwidth]{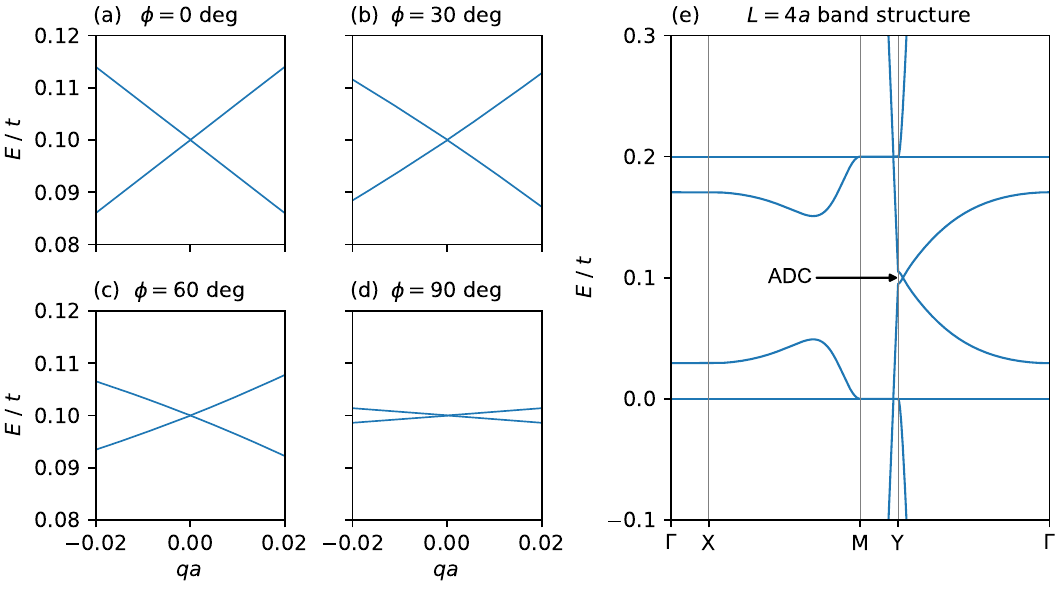}
    \caption{\textbf{Band structure around the anisotropic Dirac cone (ADC).} 
    (a-d) The band structures along small reciprocal space distances $q$ measured from the ADC crossing point at $\bm{k} = (0, \pi-V_0/2)$.
    In each subplot the angle of this cut is rotated by $\phi$ ($\phi=0$ defined as along \textbf{Y}$\rightarrow$\textbf{M}). 
    (e) The low-energy $L=4a$ Lieb SL band structure with ADC indicated. 
    }
    \label{sup_fig:around_ADC}
\end{figure}

\begin{figure}[H]
     \centering     
     \includegraphics[width=0.95\textwidth]{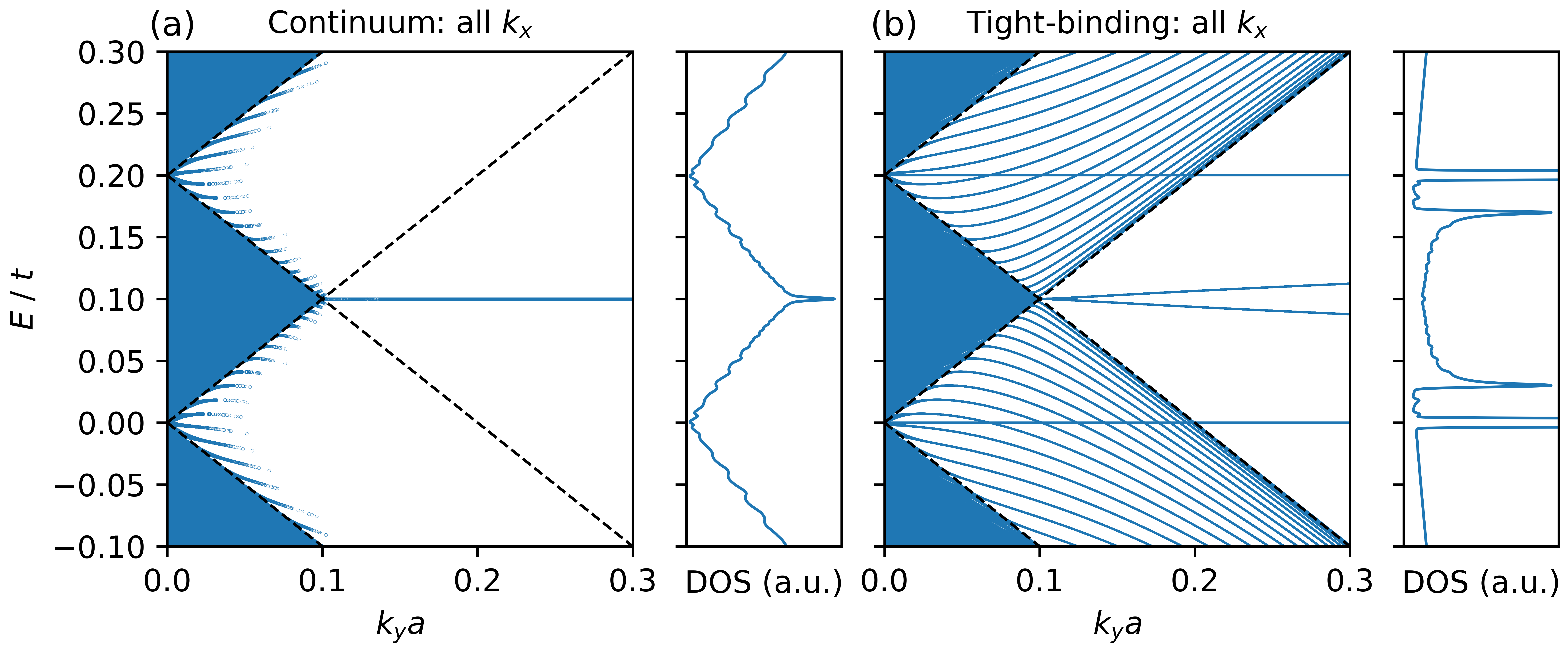}
    \caption{\textbf{Continuum vs Tight-Binding comparison for states near the Y point.} The region of the Brillouin Zone as in Fig. \ref{fig:continuum_vs_TB_figure}(a-b) but allowing for all $k_x$ values $(0 < k_x < \pi / L)$. For the continuum model description in (a), this amounts to allowing all solutions for the derived dispersion relation in Eq. \ref{sup_eqn:analytical_dispersion_relation}. In (b), we superpose the calculated bands for 20 values of $k_x$ in the range $(0 < k_x < \pi / L)$.}
    \label{sup_fig:continuum_vs_tb_all_kx}
\end{figure}

\begin{figure}[H]
     \centering
     \includegraphics[width=0.95\textwidth]{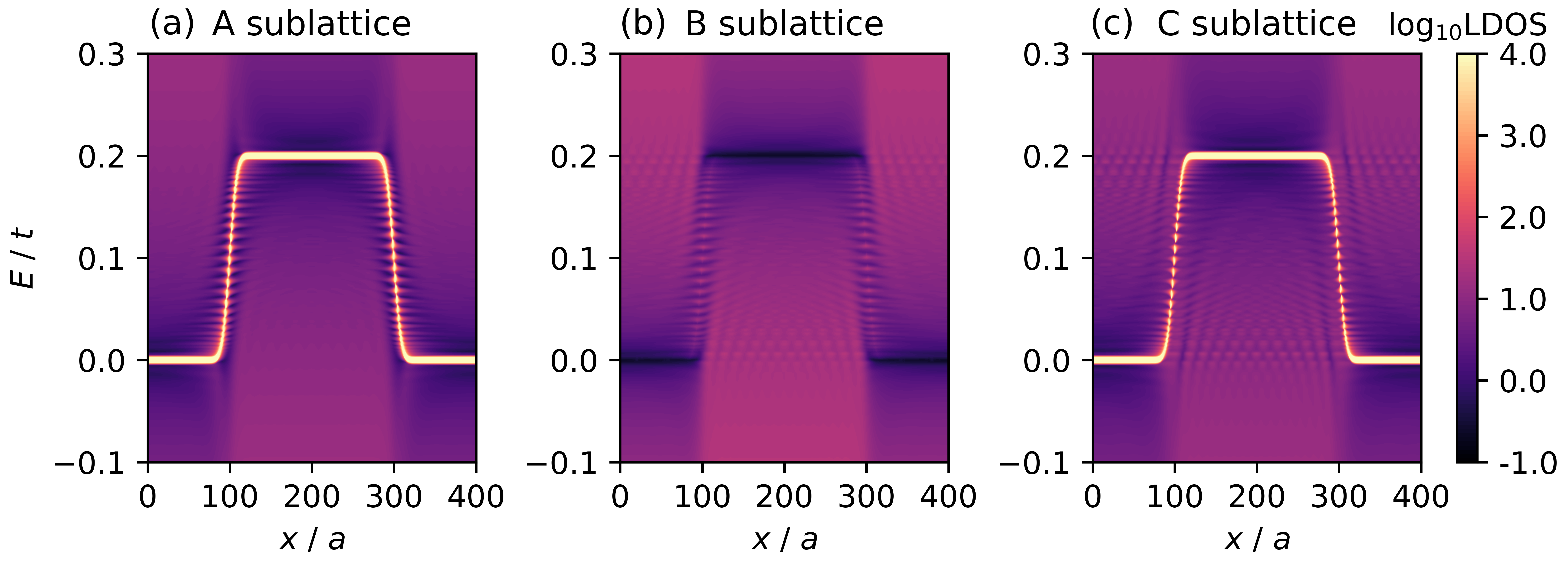}
    \caption{\textbf{LDOS spectra for A, B and C sublattices under a smoothed potential.} The potential parameters are $L=400 a$, $V_0=0.2 t$, as used in the main text in Fig. \ref{fig:smoothed_potential_figure}. The smoothness parameter $\alpha=10^{-2}$.}
    \label{sup_fig:ldos_smoothed_all_sublattices}
\end{figure}

\begin{figure}[H]
     \centering
     \includegraphics[width=0.95\textwidth]{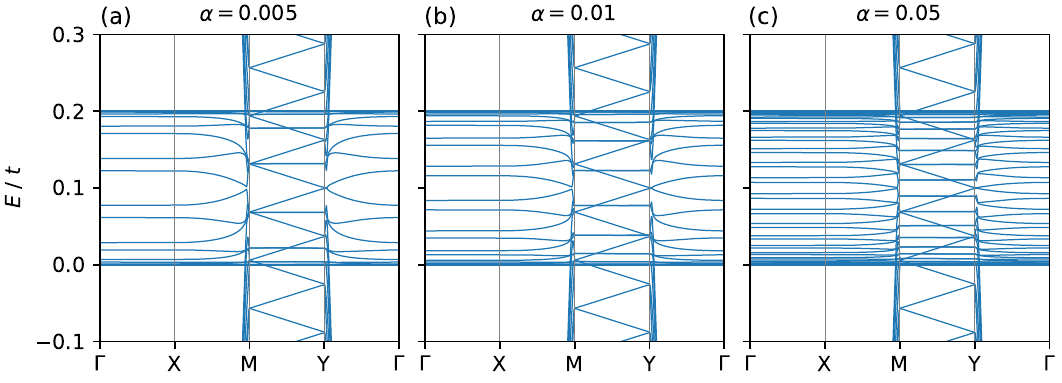}
    \caption{\textbf{Full band structures of Lieb SL for different potential smoothness.} 
    The potential parameters are $L=100 a$, $V_0=0.2 t$. The smoothness parameters are (a) $\alpha=0.005$, (b) $\alpha = 0.01$, (c) $\alpha = 0.05$. 
    The reciprocal space distances $|\bm{\Gamma}\textbf{X}|$ and $|\textbf{MY}|$ are kept artificially constant for visualisation purposes. 
    These distances are $|\bm{\Gamma}\textbf{X}| = |\textbf{MY}| = |\textbf{Y}\bm{\Gamma}| / L$ due to folding of the BZ. 
    Each pair of discrete lattice symmetry broken states generate two additional partial flat bands along $\bm{\Gamma}\textbf{X}$ and a doubly degenerate flat band along \textbf{MY}. }
    \label{sup_fig:smoothed_potential_full_bands}
\end{figure}

\begin{figure}[H]
     \centering
     \includegraphics[width=0.95\textwidth]{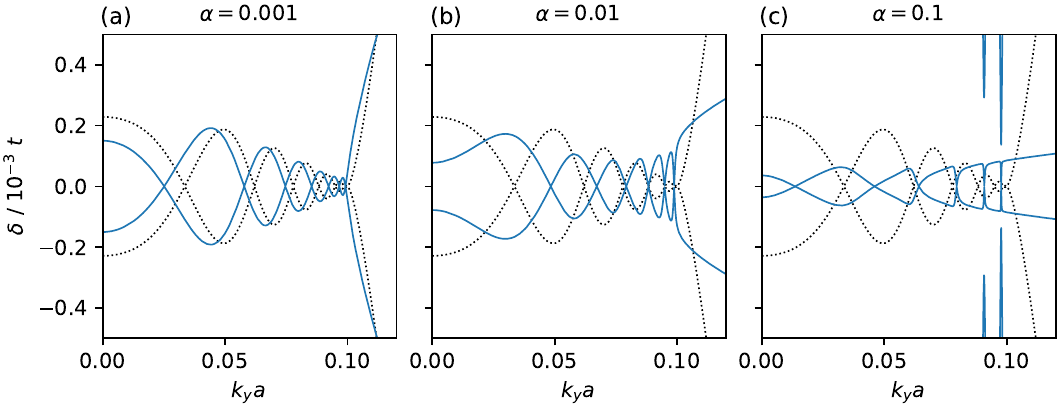}
    \caption{\textbf{The braided bands at $\bm{E=V_0/2}$ for different SL potential smoothness.} The potential parameters are $L=400 a$, $V_0=0.2 t$, as used in the main text in Fig. \ref{fig:smoothed_potential_figure}. The smoothness parameters values used are (a) $\alpha = 0.001$, (b) $\alpha = 0.01$, (c) $\alpha = 0.1$.}
    \label{sup_fig:smoothed_potential_skt_states}
\end{figure}

\begin{figure}[H]
     \centering     
     \includegraphics[scale=1]{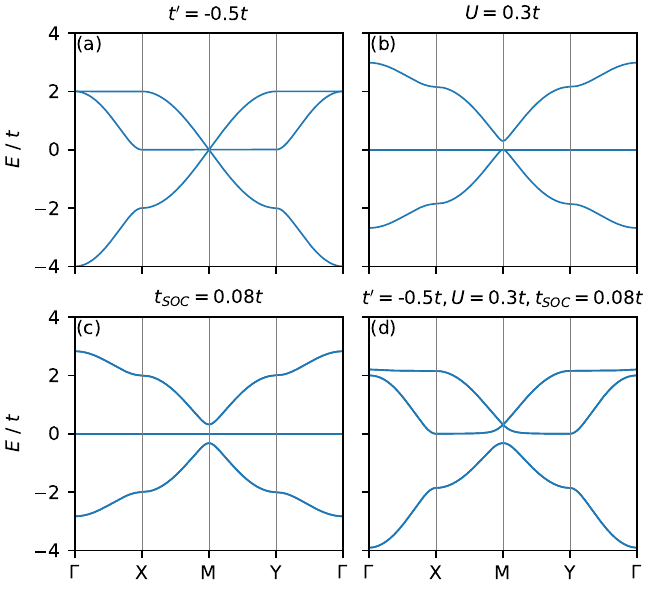}
    \caption{\textbf{Adding additional parameters to the pristine Lieb lattice.}
    Pristine Lieb lattice (absence of periodic potential) band structures upon inclusion of (a) a next-nearest neighbour interaction, (b) a mass term, (c) spin-orbit coupling, and (d) all of the above. 
    The values used are shown in the respective panels.
    High symmetry points of the pristine Lieb lattice are $\bm{\Gamma}(0,0)$, \textbf{X}$(\frac{\pi}{a}, 0)$, \textbf{M}$(\frac{\pi}{a}, \frac{\pi}{a})$, \textbf{Y}$(0, \frac{\pi}{a})$.}
    \label{sup_fig:pristine_bands_vary_params}
\end{figure}

\begin{figure}[H]
     \centering     
     \includegraphics[width=0.95\textwidth]{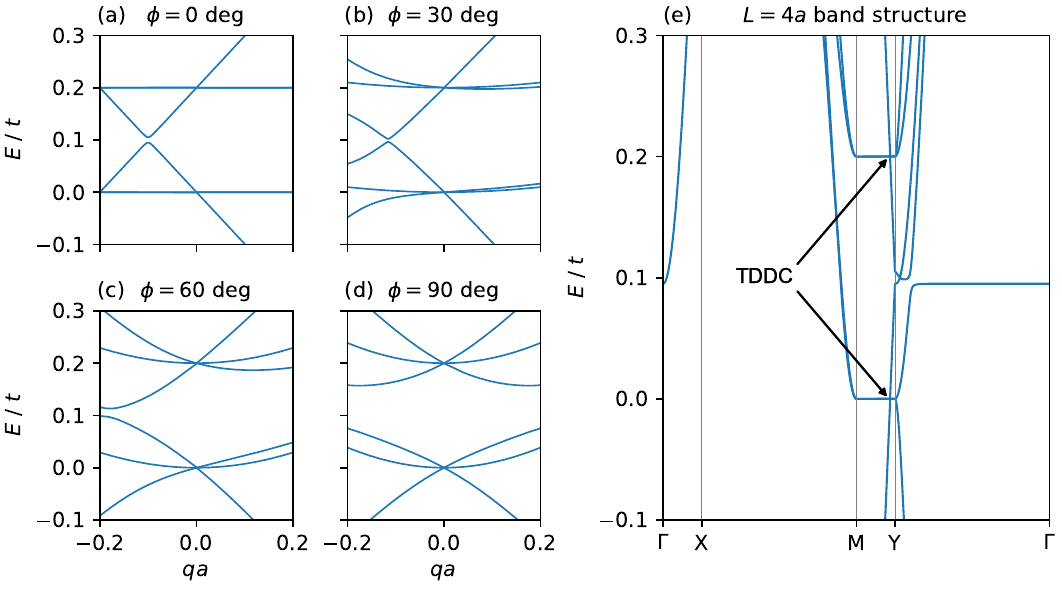}
    \caption{\textbf{Band structure around the TDDC with NNN hoppings.} The same as Figure \ref{sup_fig:around_TDDC} but with a NNN hopping term $t'=-0.5 t$. 
    The flat band disperses for all directions except for $\phi=0$ (along \textbf{Y}$\rightarrow$\textbf{M}).}
    \label{sup_fig:around_TDDC_NNN}
\end{figure}

\begin{figure}[H]
     \centering     \includegraphics[width=0.95\textwidth]{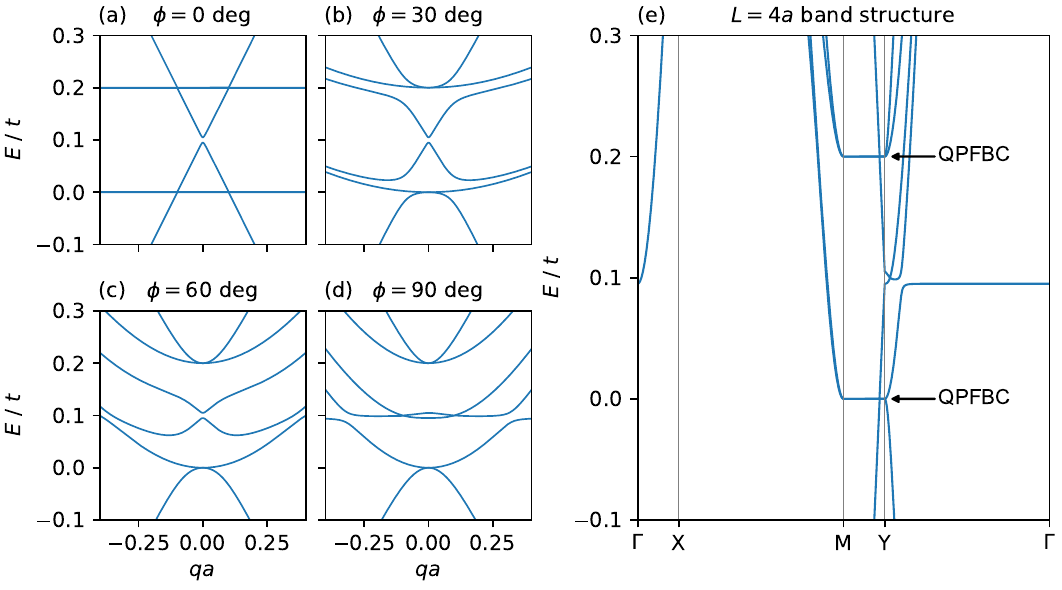}
    \caption{\textbf{Band structure around the Y point with NNN hoppings.}
    The same as Figure \ref{sup_fig:around_Y} but with a NNN hopping term $t'=-0.5 t$.
    The dispering flat band turns the quadratic flat band crossings (QFBCs) into quadratic band crossings (QBCs) with curvatures of opposite (same) sign at $E=0$ $(V_0)$.}
    \label{sup_fig:around_Y_NNN}
\end{figure}

\begin{figure}[H]
     \centering     \includegraphics[width=0.95\textwidth]{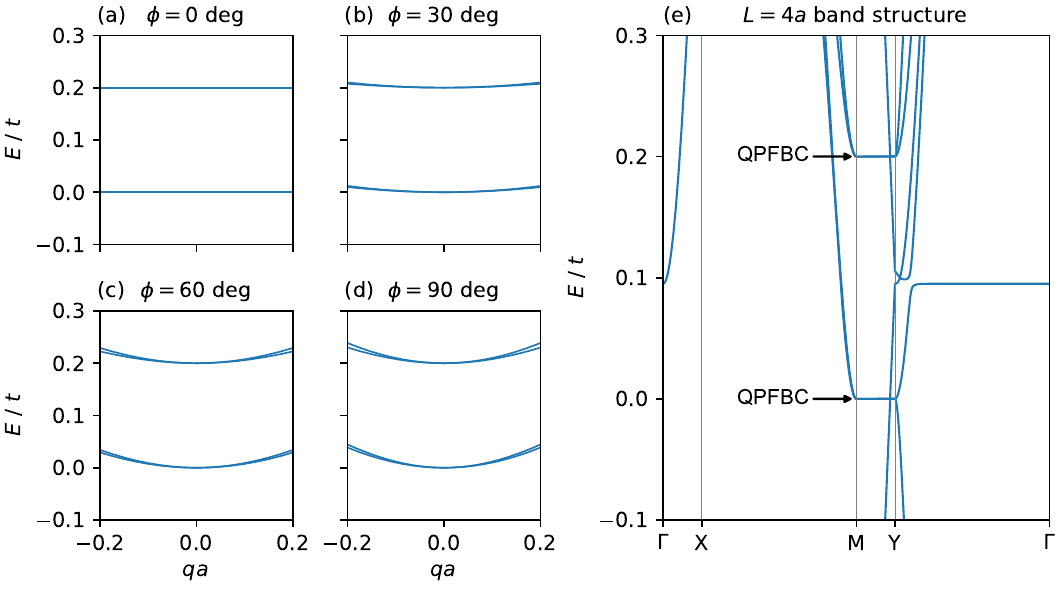}
    \caption{\textbf{Band structure around the M point with NNN hoppings.}
    The same as Figure \ref{sup_fig:around_M} but with a NNN hopping term $t'=-0.5 t$.}
    \label{sup_fig:around_M_NNN}
\end{figure}

\begin{figure}[H]
     \centering     \includegraphics[width=0.95\textwidth]{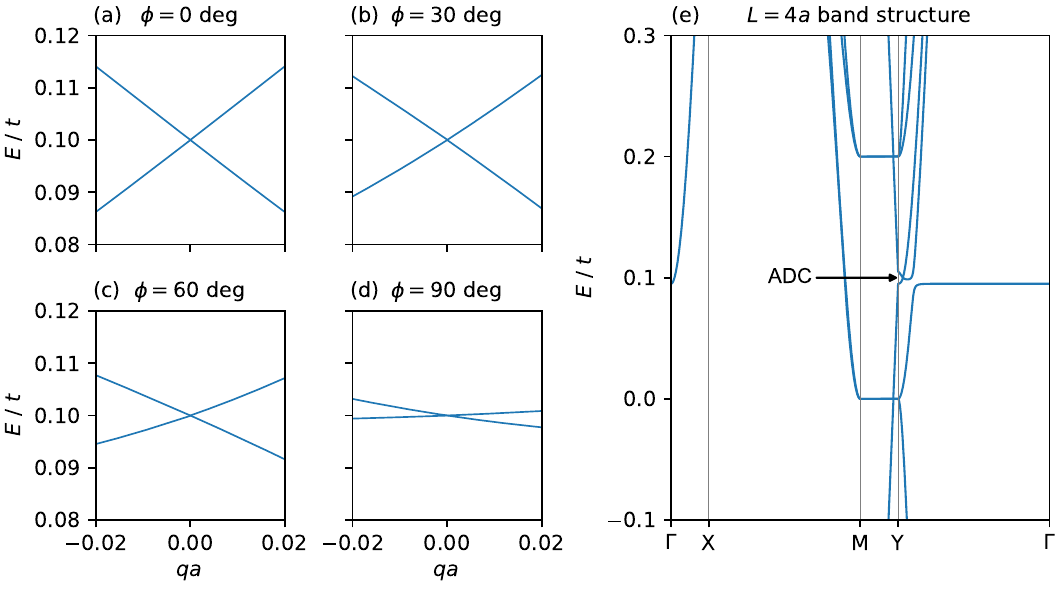}
    \caption{\textbf{Band structure around the ADC with NNN hoppings.}
    The same as Figure \ref{sup_fig:around_ADC} but with a NNN hopping term $t'=-0.5 t$.}
    \label{sup_fig:around_ADC_NNN}
\end{figure}

\begin{figure}[H]
    \centering     \includegraphics[width=0.95\textwidth]{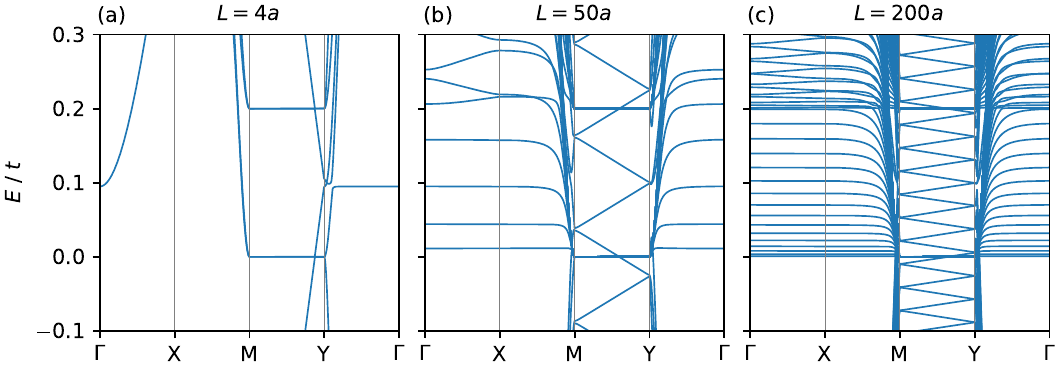}
    \caption{\textbf{Lieb SL band structure with next-nearest neighbour (NNN) hoppings.}
    Band structures along high-symmetry directions for SL periodicities (a) $L=4a$, (b) $L=50a$, and (c) $L=200a$ with a NNN hopping parameter $t'=-0.5t$.
    The reciprocal space distances $\Gamma$X and MY are kept artificially constant to visualise the band folding MY which remains unchanged from the nearest-neighbour only case. 
    }
    \label{sup_fig:full_bands_NNN_varyL}
\end{figure}

\begin{figure}[H]
    \centering     \includegraphics[width=0.95\textwidth]{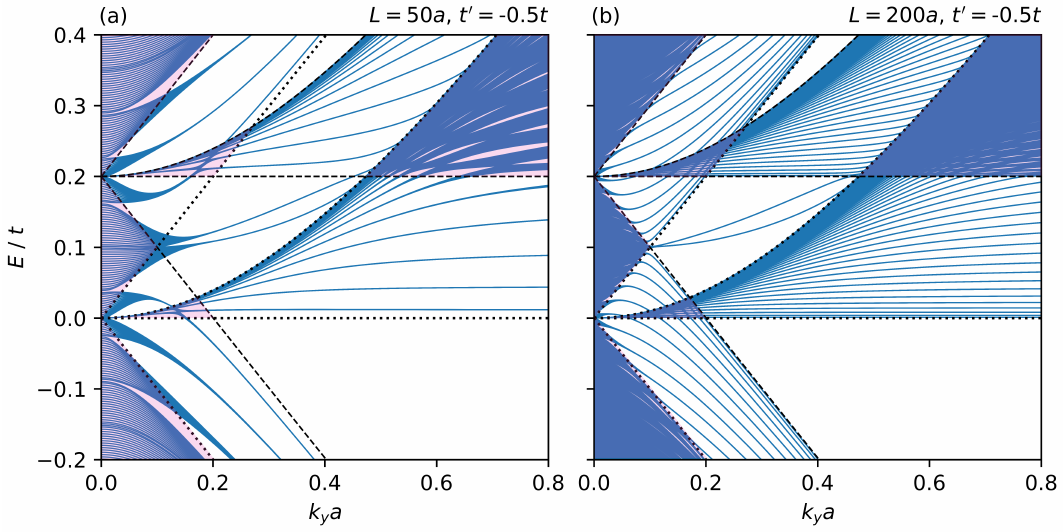}
    \caption{\textbf{Bands near the Y point for multiple $k_x$ values with next-nearest neighbours (NNN).}
    Bands near the \textbf{Y} point plotted for 20 $k_x$ values in the range $k_x: [0, \pi/L]$ with NNN hopping parameter $t'=-0.5t$ and SL periodicities (a) $L=50a$, and (b) $L=200a$. The shaded pink regions show where electronic states are allowed in both well and barrier regions according to their respective pristine dispersions.
    }
    \label{sup_fig:near_Y_NNN_vary_kx}
\end{figure}

\begin{figure}[H]
    \centering     \includegraphics[width=0.95\textwidth]{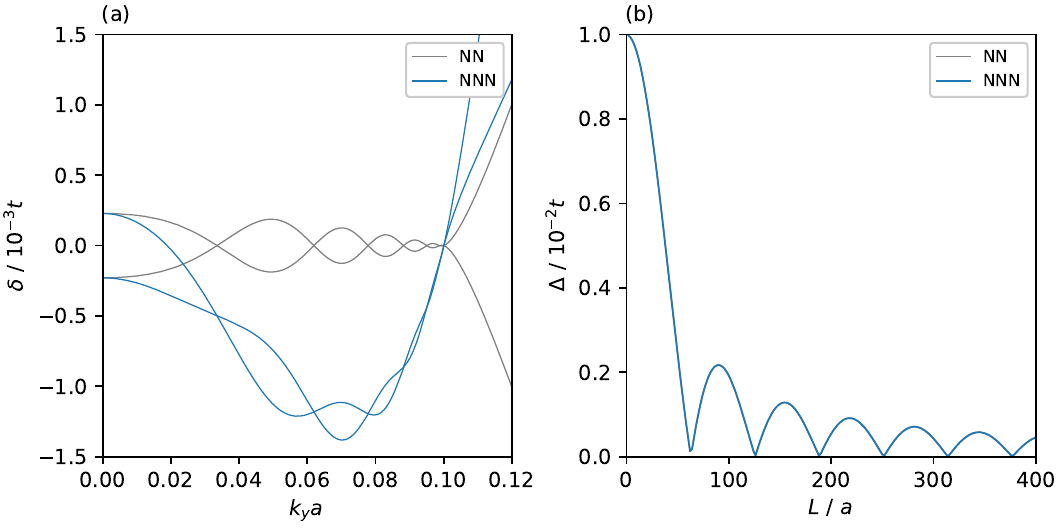}
    \caption{\textbf{States near $E=V_0/2$ with NNN.}
    (a) Inclusion of NNN (blue curve) shifts the energies of additional crossings generated at large SL periodicities compared to the NN only case (grey curve)
    (b) However, since the band folding along \textbf{MY} remains unchanged, the number and location of additional cones is identical to that of the NN case (the blue and grey curve are overlaid).
    }
    \label{sup_fig:SKT_state_NNN}
\end{figure}


\begin{figure}[H]
    \centering     \includegraphics[width=0.95\textwidth]{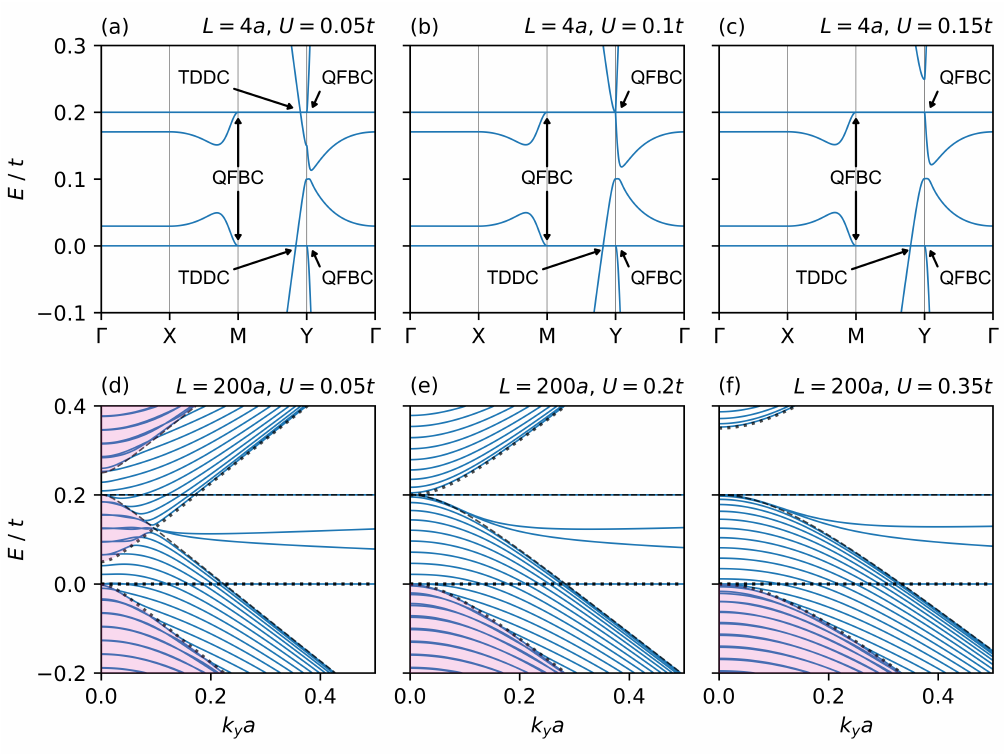}
    \caption{\textbf{Evolution of the SL dispersion and opening of a band gap with the inclusion of a mass term.} 
    Potential height is $V_0=0.2t$ in all plots. 
    In the discrete limit (top row, $L=4a$), a mass term $|U| > V_0 / 2$ is required to open a band gap. 
    In the continuum limit (bottom row, $L=200a$), a mass term $|U| > V_0$ is needed. 
    Only positive values of $U$ are shown, where the gap opens above the flat band at $E=V_0$. 
    For $U<0$ the gap appears below the flat band at $E=0$.
    The shaded pink regions show where electronic states are allowed in both well and barrier regions according to their respective pristine dispersions.
    }
    \label{sup_fig:U_limits}
\end{figure}

\begin{figure}[H]
    \centering     \includegraphics[width=0.95\textwidth]{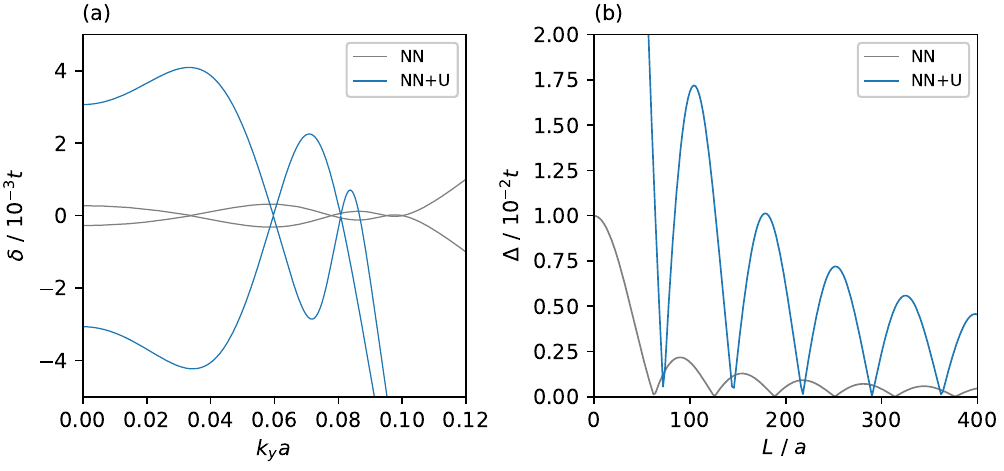}
    \caption{\textbf{States near the SKT energies with a mass term.} 
    The SL height is $V_0=0.2t$. 
    (a) Fewer additional anisotropic cones are generated at the ``SKT'' energies: $E=V_0/2$ for the NN only case, and $E=(V_0+U)/2$ when mass term is included. Here, the SL periodicity is $L=200a$. 
    (b) For a given $V_0$, a larger periodicity $L$ is required to close the gap at \textbf{Y} and generate the additional ADCs, hence fewer ADCs form in the presence of a mass term. 
    }
    \label{sup_fig:SKT_U}
\end{figure}

\begin{figure}[H]
    \centering     \includegraphics[width=0.95\textwidth]{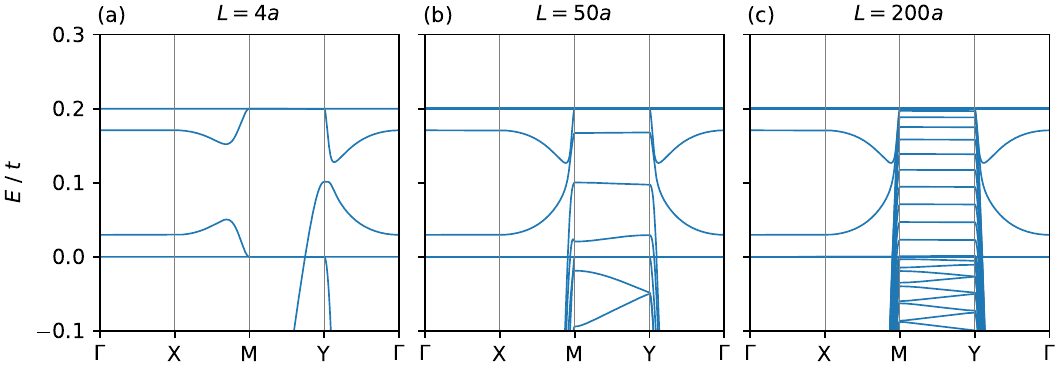}
    \caption{\textbf{Lieb SL band structure with an effective mass term.}
    Band structures along high-symmetry directions for SL periodicities (a) $L=4a$, (b) $L=50a$, and (c) $L=200a$ for a mass term $U=0.3t$. 
    The reciprocal space distances $\Gamma$X and MY are kept artificially constant to visualise the band folding MY. 
    }
    \label{sup_fig:full_bands_U_varyL}
\end{figure}

\begin{figure}[H]
    \centering     \includegraphics[width=0.95\textwidth]{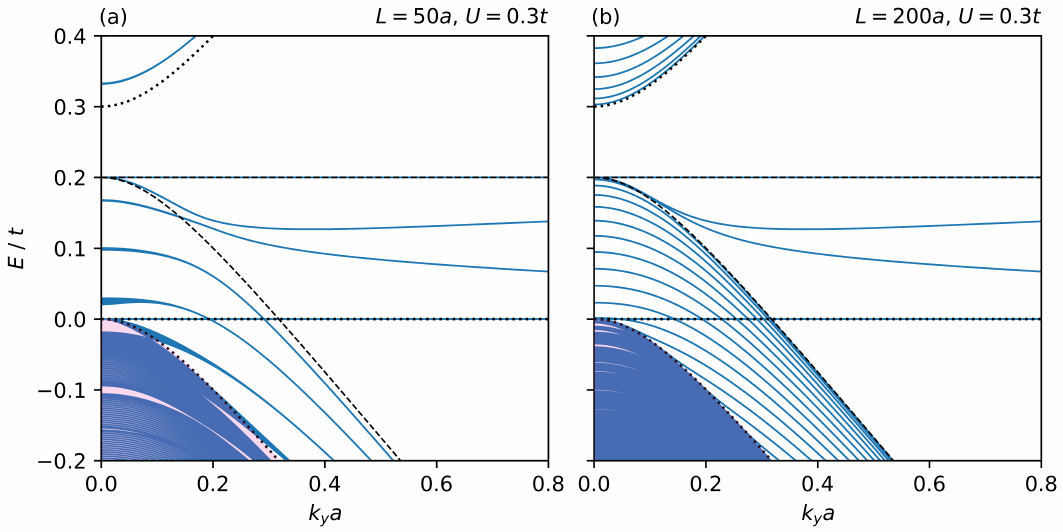}
    \caption{\textbf{Bands near the Y point for multiple $k_x$ values with an effective mass term.}
    Bands near the \textbf{Y} point plotted for 20 $k_x$ values in the range $k_x: [0, \pi/L]$ with mass term $U=0.3t$ and SL periodicities (a) $L=50a$, and (b) $L=200a$. The shaded pink regions show where electronic states are allowed in both well and barrier regions according to their respective pristine dispersions.
    }
    \label{sup_fig:near_Y_U_vary_kx}
\end{figure}


\begin{figure}[H]
    \centering     \includegraphics[width=0.95\textwidth]{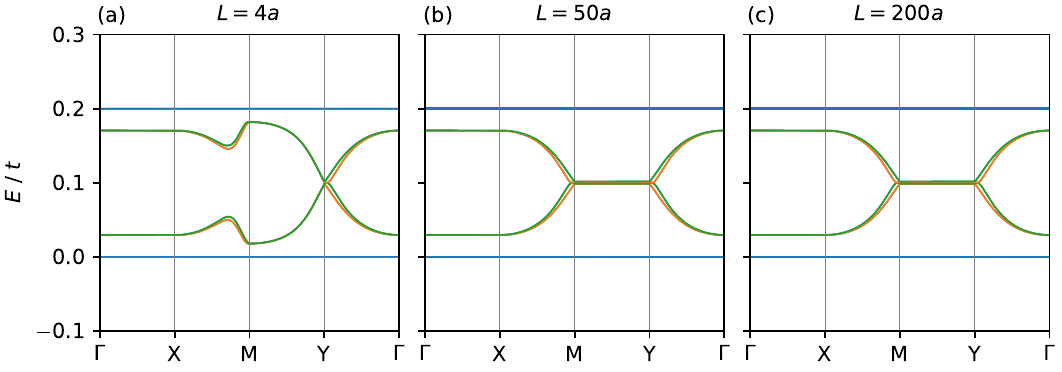}
    \caption{\textbf{Lieb SL band structure with spin-orbit coupling (SOC).}
    Band structures along high-symmetry directions for SL periodicities (a) $L=4a$, (b) $L=50a$, and (c) $L=200a$ with an SOC interaction $t_{\text{SOC}}=0.08t$. 
    The reciprocal space distances $\Gamma$X and MY are kept artificially constant to visualise the band folding MY. 
    Green and orange bands indicate the spin-splitting of the interface states that arises from discrete lattice symmetry breaking at the well-barrier interface. 
    This only occurs along \textbf{XM} and $\Gamma$Y; the spin-up and spin-down states are degenerate along $\Gamma$X and MY.
    }
    \label{sup_fig:full_bands_SOC_varyL}
\end{figure}

\begin{figure}[H]
    \centering     
    \includegraphics[width=0.95\textwidth]{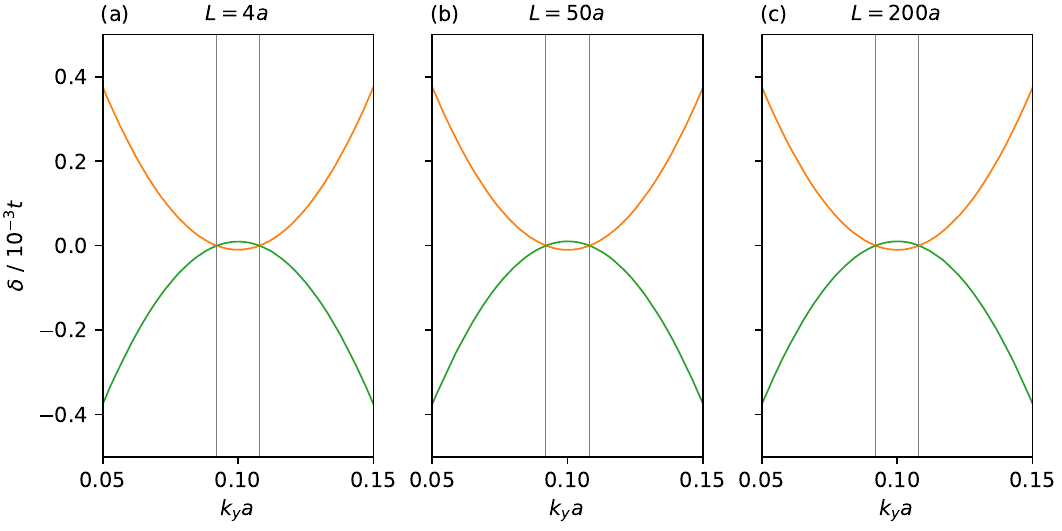}
    \caption{\textbf{States near $E=V_0/2$ with a spin-orbit coupling (SOC) term.}
    For SL periodicities (a) $L=4a$, (b) $L=50a$, (c) $L=200a$, the number and locations of the spin-polarised anisotropic cones remain unchanged, since the continuum states that would otherwise fold down to $E=V_0/2$ are prevented from doing so by the SOC term.
    }
    \label{sup_fig:SKT_state_SOC}
\end{figure}


\begin{figure}[H]
    \centering     \includegraphics[width=0.95\textwidth]{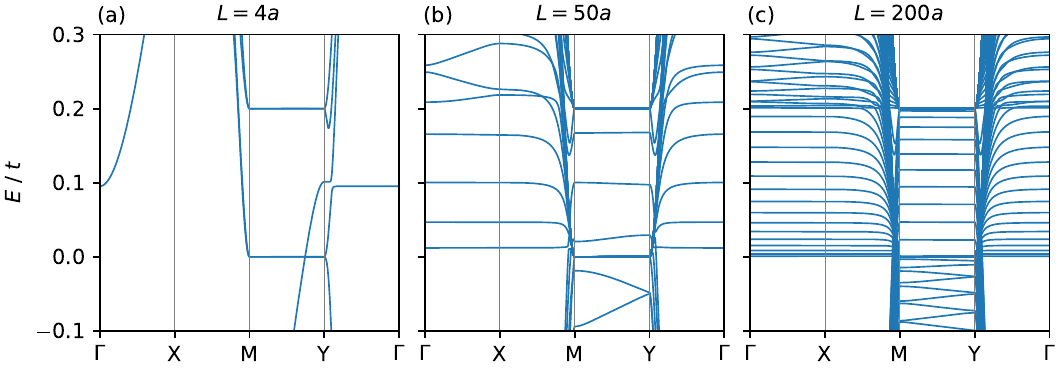}
    \caption{\textbf{Lieb SL band structure with next-nearest neighbour (NNN) hoppings and a mass term.}
    Band structures along high-symmetry directions for SL periodicities (a) $L=4a$, (b) $L=50a$, and (c) $L=200a$ with NNN hoppings $t'=-0.5t$ and mass term $U=0.3t$. 
    The reciprocal space distances $\Gamma$X and MY are kept artificially constant to visualise the band folding MY. 
    }
    \label{sup_fig:full_bands_NNN_U_varyL}
\end{figure}

\begin{figure}[H]
    \centering \includegraphics[width=0.95\textwidth]{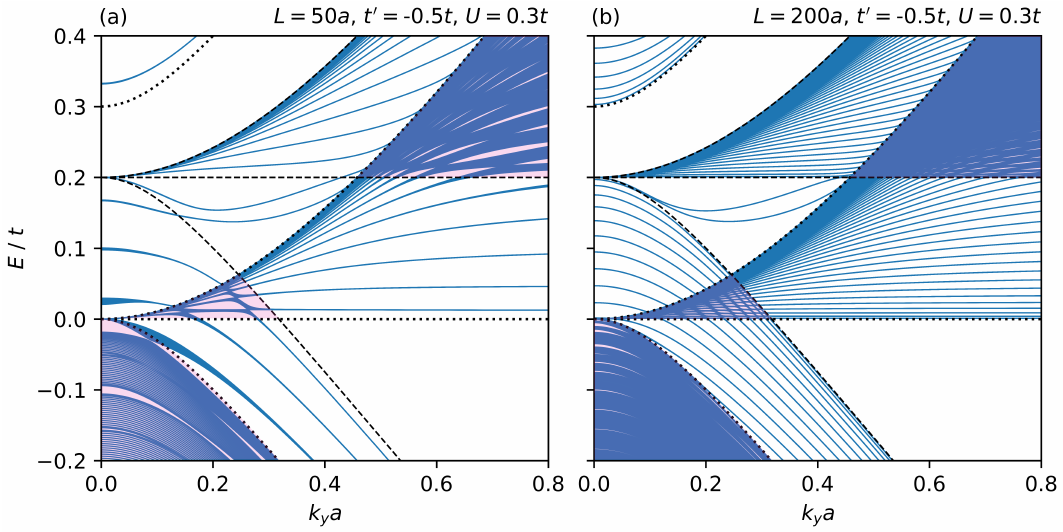}
    \caption{\textbf{Bands near the Y point for multiple $k_x$ values with next-nearest neighbour hoppings and an effective mass term.}
    Bands near the \textbf{Y} point plotted for 20 $k_x$ values in the range $k_x: [0, \pi/L]$ for SL periodicities (a) $L=50a$, and (b) $L=200a$. 
    The NNN hopping parameter $t'=-0.5t$ and mass term $U=0.3t$. The shaded pink regions show where electronic states are allowed in both well and barrier regions according to their respective pristine dispersions.
    }
    \label{sup_fig:near_Y_NNN_U_vary_kx}
\end{figure}

\begin{figure}[H]
    \centering     \includegraphics[width=0.95\textwidth]{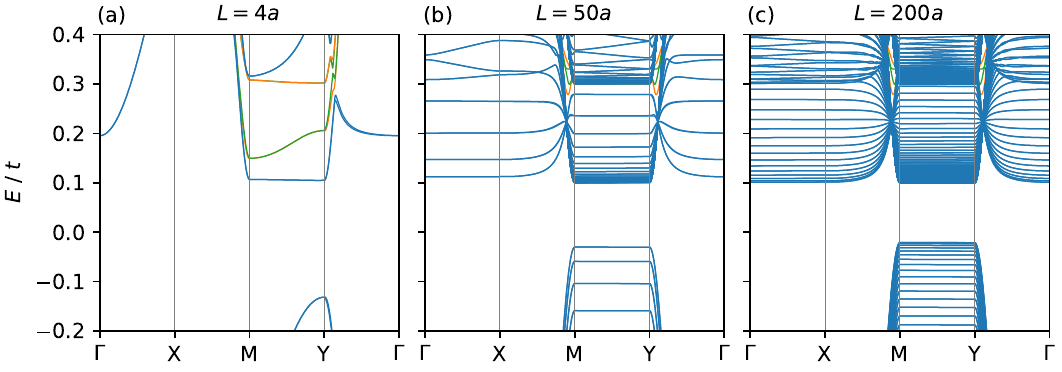}
    \caption{\textbf{Lieb SL band structure with all model parameters.}
    Band structures along high-symmetry directions for SL periodicities (a) $L=4a$, (b) $L=50a$, and (c) $L=200a$ with NNN hoppings $t'=-0.5t$, mass term $U=0.3t$, and SOC interaction $t_{\text{SOC}}=0.08t$. 
    The reciprocal space distances $\Gamma$X and MY are kept artificially constant to visualise the band folding MY. 
    }
    \label{sup_fig:full_bands_NNN_U_SOC_varyL}
\end{figure}

\begin{figure}[H]
    \centering     \includegraphics[width=0.95\textwidth]{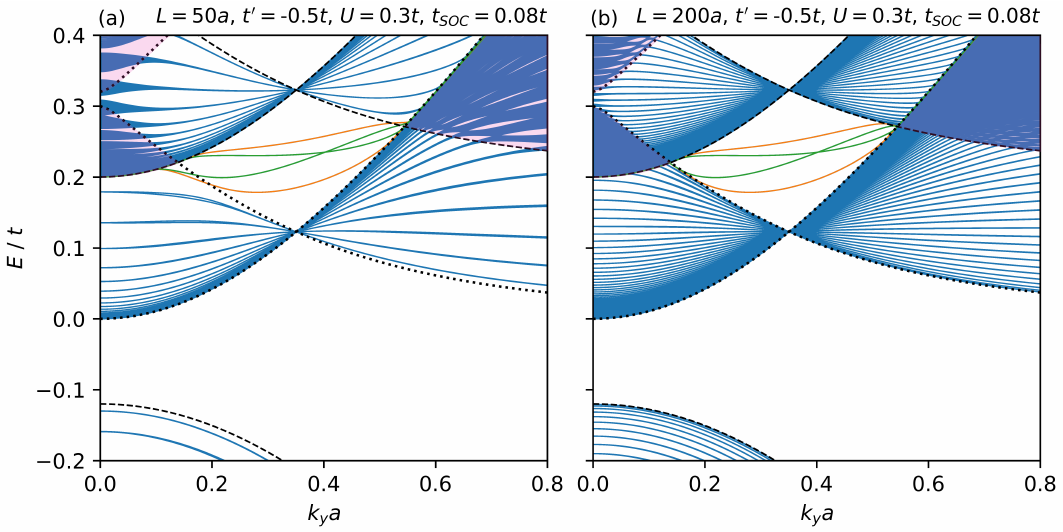}
    \caption{\textbf{Bands near the Y point for multiple $k_x$ values with all model parameters: next-nearest neighbour (NNN) hoppings, an effective mass term, and spin-orbit coupling (SOC).}
    Bands near the \textbf{Y} point plotted for 20 $k_x$ values in the range $k_x: [0, \pi/L]$ for SL periodicities (a) $L=50a$, and (b) $L=200a$. 
    The NNN hopping parameter $t'=-0.5t$, the mass term $U=0.3t$, and the SOC strength $t_{\text{SOC}}$. The shaded pink regions show where electronic states are allowed in both well and barrier regions according to their respective pristine dispersions.
    }
    \label{sup_fig:near_Y_NNN_U_SOC_vary_kx}
\end{figure}

\end{document}